\documentclass[aip,jcp,reprint,eqsecnum,superscriptaddress,twocolumn,longbibliography]{revtex4-1}
\usepackage{natbib}
\usepackage{amssymb,amsmath}

\usepackage{epsfig}
\usepackage{bm}
\usepackage{color}
\usepackage{graphicx}
\usepackage{hyperref}
\usepackage{enumerate}
\usepackage{mathpazo}

\newcommand\beq{\begin{equation}}
\newcommand\eeq{\end{equation}}
\newcommand\beqa{\begin{eqnarray}}
\newcommand\eeqa{\end{eqnarray}}

\def\bal#1\eal{\begin{align}#1\end{align}}
\newcommand{\pure}{{\text{s}}}
\def\zero{{(0)}}
\def\one{{(1)}}
\def\two{{(2)}}
\def\three{{(3)}}
\def\four{{(4)}}
\def\n{{(n)}}
\newcommand{\id}{\text{id}}
\newcommand{\ex}{\text{ex}}
\newcommand{\spl}{\text{sp}}
\newcommand{\xx}{\mathbf{x}}
\newcommand{\rr}{\mathbf{r}}

\newcommand{\pp}{\mathbf{p}}

\newcommand{\cs}{\text{CS}}
\newcommand{\FF}{\widehat{F}}
\newcommand{\Ac}{\mathcal{A}}
\newcommand{\eff}{\eta_{\text{eff}}}

\newcommand{\GG}{\mathcal{G}}

\newcommand{\lambdak}{\vartheta_1}
\newcommand{\lambdakk}{\vartheta_2}
\newcommand{\aK}{K}
\newcommand{\SE}{S}
\newcommand{\muM}{M}
\newcommand{\muMti}{m}
\newcommand{\mt}{\muMti_2}
\newcommand{\mth}{\muMti_3}
\newcommand{\mn}{\muMti_n}
\newcommand\Nc{n_c}
\newcommand{\xxx}{{(\xi)}}
\newcommand{\ZZ}{\mathcal{Z}}
\newcommand{\QQ}{\mathcal{Q}}
\newcommand{\betaT}{\beta_T}
\newcommand{\fx}{x}

\begin{document}

\title{Structural and Thermodynamic Properties of Hard-Sphere Fluids}
\author{Andr\'es Santos}
\email{andres@unex.es}
\homepage{http://www.unex.es/eweb/fisteor/andres/}
\author{Santos B. Yuste}
\email{santos@unex.es}
\homepage{http://www.unex.es/eweb/fisteor/santos/}
\affiliation{Departamento de F\'{\i}sica  and Instituto de Computaci\'on Cient\'{\i}fica Avanzada (ICCAEx), Universidad de
Extremadura, Badajoz, E-06006, Spain}
\author{Mariano L\'{o}pez de Haro}
\email{malopez@unam.mx}
\homepage{http://xml.cie.unam.mx/xml/tc/ft/mlh/}
\affiliation{Instituto de Energ\'{\i}as Renovables, Universidad Nacional Aut\'onoma de M\'exico (U.N.A.M.),
Temixco, Morelos 62580, M{e}xico}

\begin{abstract}

This Perspective article provides an overview of some of our analytical approaches to the computation of the
structural and thermodynamic properties of single-component and
multicomponent  hard-sphere fluids. For the structural
properties, they yield a thermodynamically consistent formulation,
thus improving and extending the known analytical results of the
Percus--Yevick theory. Approximate expressions linking the equation of state of the single-component fluid to the one of the multicomponent mixture are also discussed.

\end{abstract}

\maketitle

\section*{Nomenclature}

\textbf{Acronyms}

\begin{tabular}{ll}
&\\
BGHLL&Boubl\'{\i}k--Grundke--Henderson--Lee--Levesque\\
BMCSL&Boubl\'{\i}k--Mansoori--Carnahan--Starling--Leland\\
CS&Carnahan--Starling\\
DCF&direct correlation function\\
EOS&equation of state\\
FMT&fundamental measure theory\\
GMSA&generalized mean spherical approximation\\
HNC&hypernetted-chain\\
HS&hard sphere\\
LDH&linearized Debye--H\"uckel\\
MC& Monte Carlo\\
MD& molecular dynamics\\
MSA& mean spherical approximation\\
OZ&Ornstein--Zernike\\
PY&Percus--Yevick\\
RDF&radial distribution function\\
RFA&rational function approximation\\
SPT&scaled particle theory\\
\end{tabular}

\section{Introduction}
\label{sec1}

It is widely recognized that a major breakthrough in the theory of liquids was provided by the notion (already put forward by van der Waals) that in a dense fluid, the repulsive forces are mostly responsible for its structure. It is also well known that in the statistical thermodynamic approach to such a theory, there is a close connection between the thermodynamic and structural properties.\cite{BH76,M76,F85,HM06} In simple fluids, the radial distribution function (RDF) $g(r)$ (which describes the probability of finding a particle at a distance $r$ from another particle) and its close relative (through a Fourier transform),  the static structure factor $S(q)$, are the basic quantities used to discuss the structural properties. The importance of $g(r)$ arises from the fact that, given the form of the potential of the intermolecular force (which is generally assumed to be well represented by pair interactions),  the standard methods of statistical mechanics allow for the determination of all the equilibrium properties of the fluid, in particular its equation of state (EOS), if the RDF is known as a function of $r$, the number density $\rho$, and the temperature $T$.

The simplest repulsive model pair potential is that of a hard-core fluid (rods, disks, spheres, and hyperspheres) in which attractive forces are completely neglected. In fact, it is a model that has been most studied and has rendered some analytical results, although---up to this day---no general (exact) explicit expressions for the structural functions or the EOS are available, except in the one-dimensional case.  An interesting feature concerning the thermodynamic properties of such model is that the EOS depends only on the contact values of the RDF. In the absence of a completely analytical approach, the most popular methods to deal with the properties of these systems are integral equation theories and computer simulations.

In real gases and liquids at high temperatures, the  thermodynamic properties are also determined almost entirely by the repulsive forces among molecules. However,  attractive forces become significant at lower temperatures. Nevertheless, even in this case, the attractive forces affect very little the configuration of the system at moderate and high densities. These facts are taken into account in the application of the perturbation theory of fluids,\cite{S13} where hard-core fluids are used as the reference systems in the computation of the thermodynamic and structural properties of real fluids. In any case, successful results using perturbation theory are rather limited because, as mentioned above, there are in general no exact (analytical) expressions for the thermodynamic and structural properties of the reference systems, which are, in principle, required in the calculations. On the other hand, in the realm of soft condensed matter, the use of the hard-sphere (HS) model in connection with sterically stabilized colloidal systems, is quite common. This reflects the fact that presently, it is possible to prepare (almost) monodisperse spherical colloidal particles with short-ranged harshly repulsive interparticle forces that may be well described theoretically with the HS potential.

This paper presents an overview of the efforts we have made over the years to compute the thermodynamic and structural properties of hard-core systems in $d$ dimensions using relatively simple (approximate) analytical methods. Due to their particular relevance and for the sake of concreteness, we will concentrate here on  three-dimensional systems, namely, the HS fluid and the multicomponent HS fluid mixture. The paper is structured as follows. In Section \ref{sec2}, we  begin by recalling the main statistical--mechanical relationships, valid for general fluids, related  to both their structural and thermodynamic properties, as well as some key approximate results derived for them in our systems of interest; special attention is paid to the thermodynamic routes and to
a summary of results pertaining to the phase behavior of these systems, including some concerning the demixing transition.
This is followed in Sec.\ \ref{sec3} by a detailed account of an alternative methodology to the usual integral equation approach of liquid-state theory to obtain analytical results for the structural properties of HS fluids and multicomponent HS mixtures. Section \ref{sec4} is devoted to describe the routes we have followed to derive the EOS of a multicomponent HS mixture once the EOS of the monocomponent HS fluid is known, which allows one, in principle,    to probe the metastable fluid branch of the single-component fluid.  The paper is closed in Sec.\ \ref{sec6} with some concluding remarks.

\section{Statistical mechanics, thermodynamics, and structure of fluids}
\label{sec2}
\subsection{General background}
\subsubsection{One-component systems}
We begin with  the partition function $\ZZ_N(\betaT,V)$ in the canonical ensemble for a closed system of $N$ identical particles of mass $m$ enclosed in a volume $V$ at a temperature $T$ (with $\betaT\equiv {1}/{k_B T}$, $k_B$ being the Boltzmann constant), namely,
 \begin{equation}
 \label{CanZ}
  \ZZ_N(\betaT,V)=\frac{1}{N! h^{3N}}\int d \mathbf{x}^N\, e^{-\betaT H_N(\mathbf{x}^N)},
  \end{equation}
where $h$ is the Planck constant, $H_N(\mathbf{x}^N)$ the Hamiltonian of the system, $\xx^N\equiv\{\rr^N,\pp^N\}$, $d  \xx^N\equiv d  \rr^N d \pp^N$, $ \rr^N=\{\rr_1,\rr_2,\ldots,\rr_N\}$, $d  \rr^N\equiv d \rr_1d \rr_2 \cdots d \rr_N$, $ \pp^N\equiv\{\pp_1,\pp_2,\ldots,\pp_N\}$, and $ d  \pp^N\equiv d \pp_1d \pp_2 \cdots d \pp_N$ (with $\rr_\alpha$ and $\pp_\alpha$, $\alpha=1,2, \ldots, N$, denoting the position and momentum vectors of particle $\alpha$, respectively). In the general case of interacting particles, the Hamiltonian is given by $H_N(\xx^N)= H_N^\id(\mathbf{p}^N)+\Phi_N(\rr^N)$, where $H_N^\id(\mathbf{p}^N)=\sum_{\alpha=1}^N{p_\alpha^2}/{2m}$  accounts for the kinetic energy of the particles and $\Phi_N(\rr^N)$ is the intermolecular potential. Hence, one may rewrite the partition function  as $\ZZ_N(\betaT,V)=\ZZ_N^\id(\betaT,V)\QQ_N(\betaT,V)$, where
$\ZZ_N^\id(\betaT,V)={V^N}/{N!\Lambda^{3N} }$, $\Lambda=h \sqrt{{\betaT}/{2\pi m}}$ being the thermal de Broglie wavelength, and $\QQ_N(\betaT,V)=V^{-N}\int d \rr^N\, e^{-\betaT\Phi_N(\rr^N)}$ is the configuration integral.

Given the fact that the partition function and the Helmholtz free energy of the system are linked by $F(N,V,T)=-\betaT^{-1} \ln \ZZ_N(\betaT,V)$, and since the average energy, the pressure, the isothermal compressibility, and the chemical potential of the fluid are obtained from the Helmholtz free energy as
\begin{subequations}
\label{EpkappamufromF}
\begin{equation}
\label{EfromF}
\langle E\rangle=\frac{\partial (\betaT F)}{\partial \betaT},
\end{equation}
\begin{equation}
\label{pfromF}
p=-\frac{\partial F}{\partial V},
\end{equation}
\begin{equation}
\label{kappafromF}
\kappa_T^{-1}\equiv-V\frac{\partial p}{\partial V}=V \frac{\partial^2 F}{\partial V^2},
\end{equation}
\begin{equation}
\label{mufromF}
\mu=\frac{\partial F}{\partial N},
\end{equation}
\end{subequations}
respectively, in order to derive such thermodynamic properties all that one, in principle, needs is the explicit form of $\Phi_N(\rr^N)$ and the subsequent computation of the configuration integral. However, in practical terms, dealing with an $N$-particle problem when $N\sim 10^{23}$ is not, in general, feasible. Nevertheless, one way to attempt to make it feasible is by reducing the problem of a macroscopic fluid in a volume $V$ to a sum of an increasing number of tractable isolated few ($n=1$, $2$, $3$, \ldots) particle problems, where each group of particles moves alone in the volume $V$ of the system. In the case of the pressure, this approach leads formally to the virial expansion of the EOS, which reads
  \begin{equation}
  \label{pressure}
  \betaT p(\rho, T)= \rho  + \sum_{n=2}^\infty B_n(T) \rho^{n},
  \end{equation}
where $\rho=N/V$ is the number density. Such an expansion was introduced empirically by Thiesen\cite{T85} and (independently) by Kamerlingh Onnes\cite{KO01} with the aim of providing a mathematical representation of experimental results of the dependence of pressure on temperature and density of gases and liquids through an expansion of the pressure in powers of density. In fact, it was Kamerlingh Onnes who named the coefficients $B_n(T)$ in the expansion the virial coefficients. Once the statistical--mechanical rigorous derivation of this series became available, it was immediate to relate the virial coefficients  to intermolecular interactions involving $n$ particles. Such relationships, derived originally by Mayer and Mayer,\cite{MGM40} imply that the $n$th virial coefficient can be written as a sum of integrals represented by the so-called $n$-particle star graphs. It is unfortunate, however, that---in general---the actual computation of the virial coefficients is a formidable task, and that the radius of convergence of the series in Eq.\ (\ref{pressure}) is not known. Nevertheles, if $\rho$ is small enough and if the required first few virial coefficients are available, a reasonable approximation to the EOS of the system may thus be derived. We will come back to this point later on for the specific case of HS fluids and fluid  mixtures. A review on virial expansions, including an extensive list of references and a description of the difficulties associated with the computation of higher virial coefficients, has been written by Masters.\cite{M08a} In a rather recent paper, Hoover and Hoover\cite{HH20} have provided a nice account on the early history of the numerical computation of virial coefficients and the more recent developments.

Now we turn to the structural properties. Although of course it is the full $N$-body probability distribution function $\varrho_N(\mathbf{x}^N)$ that contains all the statistical--mechanical information about the system, marginal few-body distributions may be enough for the most relevant quantities. Let us now introduce the reduced $s$-body  correlation functions $f_s(\xx^s)$ so that $f_s(\xx^s)d \xx^s$ corresponds to the number of groups of $s$ particles such that one particle lies inside a volume $d \xx_1$ around the (one-body) phase-space point $\xx_1$, other particle lies inside a volume $d \xx_2$ around the (one-body) phase-space point $\xx_2$, \ldots, and so on. Hence,
 \begin{equation}
f_s(\mathbf{x}^s)=
\frac{N!}{(N-s)!}\int d \mathbf{x}_{s+1}\int d \mathbf{x}_{s+2}\cdots \int d \mathbf{x}_{N}\,
\varrho_N(\mathbf{x}^N).
\end{equation}

Since, in equilibrium, the momenta of all the particles are uncorrelated, it is also convenient to introduce the configurational $s$-body  correlation functions $n_s(\rr^s)$ obtained by integrating $f_s(\xx^s)$ over the momenta. These correlation functions are translationally invariant. In the particular case $s=2$, $n_2(\mathbf{r}_1, \mathbf{r}_2)=\int d \mathbf{p}_{1}\int d \mathbf{p}_{2}\,f_2(\mathbf{x}_1, \mathbf{x}_2)$, and it follows from the constant number of particles that $\int d \mathbf{r}_{1}\int d \mathbf{r}_{2}\,n_2(\mathbf{r}_1, \mathbf{r}_2)=N(N-1)$.

Recalling that in the canonical ensemble $\varrho_N(\mathbf{x}^N)$ is given by
 \begin{equation}
  \varrho_N(\mathbf{x}^N)=\frac{e^{-\betaT H_N(\mathbf{x}^N)}}{N! h^{3N}\ZZ_N^\id \QQ_N},
    \end{equation}
one has
\begin{equation}
 n_2(\mathbf{r}_1,  \mathbf{r}_2)=\frac{N(N-1)}{V^N \QQ_N}\int d \mathbf{r}_{3}\cdots \int d \mathbf{r}_{N}\,e^{-\betaT \Phi_N(\mathbf{r}^N)}.
  \end{equation}
In turn, defining the \emph{pair} correlation function $g_2(\mathbf{r}_1, \mathbf{r}_2)$ by $n_2(\mathbf{r}_1, \mathbf{r}_2)=\rho^2 {g_2(\mathbf{r}_1, \mathbf{r}_2)}$, one obtains
\begin{equation}
g_2(\mathbf{r}_1, \mathbf{r}_2)=\frac{V^{-(N-2)}}{\QQ_N}\int d \rr_3\cdots\int d \rr_N\, e^{-\betaT\Phi_N(\rr^N)},
\end{equation}
where we have taken into account that $N-1\approx N$.

A fluid is also rotationally invariant. Therefore, assuming central forces and using  translational invariance, it follows that $g_2(\mathbf{r}_1, \mathbf{r}_2)=g_2(\mathbf{r}_1-\mathbf{r}_2)=g(r)$ with $r\equiv |\mathbf{r}_1-\mathbf{r}_2|$, so that this pair correlation function is precisely the RDF. Note that, once again, knowledge of the intermolecular potential would be, in principle, enough to obtain $g(r)$. Also note that, if a given particle is taken to be at the origin,  then the \emph{local} average density at a distance $r$  from that particle is $\rho g(r)$. Therefore, as already mentioned, $g(r)$  is a measure of the probability of finding a particle at a distance $r$ away from a given reference particle, \emph{relative} to that for an ideal gas.

Other important related structural quantities are the total correlation function $h(r)\equiv g(r)-1$, the static structure factor
\begin{equation}
\label{S(q)bis}
S(q)=1 + \rho \int d\mathbf{r}\, h(r)e^{-\imath \mathbf{q}\cdot \mathbf{r}}
 \end{equation}
(where $\imath$ is the imaginary unit), and the direct correlation function (DCF) $c(r)$, a quantity that we will come back to later. The relevance of these structural quantities also resides in the fact that they may be related to the thermodynamic properties of the fluid.\cite{S16} In particular, restricting ourselves to pairwise additive intermolecular potentials, namely,
\begin{equation}
\label{pairwise}
\Phi_N(\mathbf{r}^N)=\sum_{\alpha=1}^{N-1}\sum_{\beta=\alpha+1}^N\phi(r_{\alpha\beta})=\frac{1}{2}\sum_{\alpha\neq \beta}\phi(r_{\alpha\beta}),
\end{equation}
it follows that the average energy of the system $\langle E\rangle$ is given by
\begin{equation}
\label{averageE}
\frac{\langle E\rangle}{N}=\frac{3}{2}k_B T+\frac{\rho}{2}\int d \mathbf{r}\, \phi(r) g(r),
\end{equation}
 where the first term on the right-hand side accounts for the kinetic energy and the second one for the potential energy. On the other hand, the compressibility factor of the fluid is given by
\begin{equation}
\label{virpres}
Z\equiv \frac{\betaT p}{\rho}=1-\frac{\betaT\rho }{6}\int d \mathbf{r}\, r \frac{d \phi(r)}{d  r} g(r).
\end{equation}

Two other important thermodynamic quantities are the reduced isothermal compressibility (henceforth referred to as isothermal susceptibility) $\chi\equiv \rho \kappa_T/\betaT$  and the chemical potential $\mu$. For the former, the connection with the structural properties is through
\begin{equation}
\label{isocomp}
\chi=1+ \rho \int d \mathbf{r}\, h(r)=S(0).
\end{equation}

In the case of the chemical potential, what one does is to consider an $(N+1)$-particle system such that the new intermolecular potential is
$\Phi_{N+1}^\xxx(\rr^{N+1})=\sum_{\alpha=1}^{N-1}\sum_{\beta=\alpha+1}^N \phi(r_{\alpha\beta})+\sum_{\beta=1}^N\phi_{\text{test}}^\xxx(r_{0\beta})$,  where the additional (test) particle has been labeled as particle $\alpha=0$ and a continuous \emph{coupling parameter} $\xi$, such that its value $0\leq\xi\leq 1$ controls the strength of the interaction of the test particle to the rest of particles, is introduced. The boundary values of $\phi_{\text{test}}^\xxx(r)$ and of the corresponding RDF,  $g_{\text{test}}^\xxx(r)$, are
\begin{equation}
\label{coupling}
\{\phi_{\text{test}}^\xxx(r),g_{\text{test}}^\xxx(r)\}=\begin{cases}
 \{ 0,1\},&\xi=0,\\
  \{\phi(r),g(r)\},&\xi=1.
  \end{cases}
  \end{equation}
The connection between the chemical potential  and the structure of the fluid turns out to be given by\cite{S16,S12b,SR13}
\begin{subequations}
\begin{equation}
\betaT\mu=\ln\left(\rho\Lambda^3\right)+\betaT \mu^\ex,
\end{equation}
\begin{equation}
\label{chempot}
\mu^\ex=\rho \int_0^1 d \xi \int d \rr\, g_{\text{test}}^\xxx(r)\frac{\partial \phi_{\text{test}}^\xxx(r)}{\partial\xi}.
\end{equation}
\end{subequations}

The above relationships between structural properties and thermodynamic properties reflect the importance of the RDF in the theory of liquids. Since these thermodynamic properties are also linked to partial derivatives of the Helmholtz free energy of the fluid, it is usual to refer to Eqs.\ (\ref{averageE}), (\ref{virpres}), (\ref{isocomp}), and (\ref{chempot}) as the energy route, the virial route, the compressibility route, and the chemical-potential ($\mu$) route to the EOS, respectively.

It is also possible to derive a more direct free-energy route by a procedure analogous to the one used in the $\mu$ route, except that  now, the charging process is parameterized by a common coupling parameter $\xi$ that affects all the particles of the system, not only a test particle. The associated pair potential, $\phi^\xxx(r)$, and RDF, $g^\xxx(r)$,  satisfy  the same boundary conditions as in Eq.\ \eqref{coupling}, namely,
\begin{equation}
\label{couplingF}
\{\phi^\xxx(r),g^\xxx(r)\}=\begin{cases}
 \{ 0,1\},&\xi=0,\\
  \{\phi(r),g(r)\},&\xi=1.
  \end{cases}
  \end{equation}
Thus, the free-energy route becomes\cite{S16}
\begin{subequations}
\begin{equation}
\betaT a\equiv\frac{\betaT F}{N}=\ln\left(\rho\Lambda^3\right)-1+{\betaT a^\ex},
\end{equation}
\begin{equation}
\label{freeenergy}
{a^\ex}=\frac{\rho}{2} \int_0^1 d \xi \int d \rr\, g^\xxx(r)\frac{\partial \phi^\xxx(r)}{\partial\xi},
\end{equation}
\end{subequations}
where we have introduced the Helmholtz free energy per particle ($a$) and its excess part ($a^\ex$).

Equation \eqref{freeenergy} can be termed as a \emph{master route} in the sense that it encompasses the energy and virial routes as particular choices of the charging process.\cite{S16} The choice $\phi^\xxx(r)=\xi \phi(r)$ is not but an energy rescaling yielding $g^\xxx(r;\betaT)=g(r;\xi\betaT)$, so that combining Eqs.\ \eqref{EfromF} and \eqref{freeenergy} one recovers the energy route [Eq.\ \eqref{averageE}].
Alternatively, the choice $\phi^\xxx(r)=\phi(r/\xi)$ is a distance rescaling yielding $g^\xxx(r;\rho)=g(r;\rho\xi^3)$. Next, the thermodynamic relation [see Eq.\ \eqref{pfromF}]
\begin{equation}
\label{Zfroma}
Z=1+\rho\frac{\partial(\betaT a^\ex)}{\partial\rho}
\end{equation}
gives Eq.\ \eqref{virpres} from Eq.\ \eqref{freeenergy}.

Now we return to the DCF $c(r)$. This structural quantity was introduced in 1914 through the Ornstein--Zernike (OZ) relation,  which simply states that the total correlation function between two particles is the sum of the direct correlation between them and the correlation mediated by the other particles, namely,
  \begin{equation}
  \label{OZ}
  h(r_{12})=c(r_{12})+\rho\int \text{d}\mathbf{r}_3\, {c(r_{13})}h(r_{23}).
  \end{equation}
According to this, it is expected that {(at least for short-ranged potentials)} the range of $h(r)$ is greater than that of $c(r)$,  which in turn should be approximately  equal to the range of $\phi(r)$. Furthermore, in Fourier space it follows from Eq.\ (\ref{OZ}) that
\begin{equation}
\label{OZFourier}
\widetilde{h}(q)=\frac{\widetilde{c}(q)}{1-\rho\widetilde{c}(q)},\quad \widetilde{c}(q)=\frac{\widetilde{h}(q)}{1+\rho \widetilde{h}(q)},
 \end{equation}
where $\widetilde{h}(q)$ and $\widetilde{c}(q)$ are the Fourier transforms of $h(r)$ and $c(r)$, respectively.
Hence, one may re-express the compressibility route to the EOS [Eq.\ \eqref{isocomp}]  as
   \begin{equation}
   \label{compres2}
\chi=\frac{1}{1-\rho \widetilde{c}(0)}.
\end{equation}

Another relevant structural quantity is the cavity (or background) function  $y(r)\equiv g(r)e^{\betaT\phi(r)}$, so that $y(r)=g(r)$ outside the range of the potential. The cavity function is much more regular than the RDF. In fact, it is continuous even if the interaction potential is discontinuous or diverges. For completeness, we now also introduce the indirect correlation function as $\gamma(r)\equiv h(r)-c(r)$.

Note that Eq.\ (\ref{OZ}) defines $c(r)$ but is not a closed equation.  Exact statistical mechanics\cite{HM06} allows one to write the following relationship
\begin{equation}
\label{Closureeq}
c(r)=h(r)-\ln [1+h(r)]-\betaT \phi(r) + b(r),
\end{equation}
where the function $b(r)$, named bridge function after its diagrammatic characterization, is a functional of the total correlation function, i.e., its value at distance $r$ depends on the values of $h(r)$ at all distances. Unfortunately this bridge function is not exactly known and in order to get some progress one must resort to approximations.

\subsubsection{Multicomponent systems}
The structural properties and their relationship with thermodynamic quantities may also be considered for  mixtures of $\Nc$ components. Let $N_i$ be the number of particles of species $i$ in the mixture (so that the total number of particles  is $N =\sum_{i=1}^{\Nc} N_i$). In turn, the mole fraction of species $i$ is $x_i=N_i/N$, with $\sum_{i=1}^{\Nc} x_i=1$, while the number density of species $i$ is $\rho_i=N_i/V=\rho x_i$. Further, the interaction potential between a particle of species $i$ and a particle of species $j$ is denoted by  $\phi_{ij}(r)$. In this system it is also convenient to introduce at this stage the  RDF for the pair  of particles of species $i$ and $j$ as $g_{ij}(r)$. The associated total correlation function and cavity function are thus
$h_{ij}(r)=g_{ij}(r)-1$ and $y_{ij}(r)=g_{ij}(r)e^{\betaT \phi_{ij}(r)}$, respectively.
The OZ equation for the multicomponent system reads
\begin{equation}
 \label{OZMulti}
  h_{ij}(r_{12})=c_{ij}(r_{12})+\rho\sum_{k=1}^{\Nc} x_k\int d \mathbf{r}_3\, c_{ik}(r_{13})h_{jk}(r_{23}),
  \end{equation}
which serves as a definition of the DCF $c_{ij}(r)$. In Fourier space and in matrix form, Eq.\ \eqref{OZMulti} becomes
\begin{equation}
\label{OZMulti2}
\widehat{\mathsf{h}}(q)=\left[\mathsf{I}-\widehat{\mathsf{c}}(q)\right]^{-1}-\mathsf{I},\quad \widehat{\mathsf{c}}(q)=\mathsf{I}-\left[\mathsf{I}+\widehat{\mathsf{h}}(q)\right]^{-1},
\end{equation}
where $\mathsf{I}$ is the $\Nc\times \Nc$ identity matrix, while the elements of the matrices $\widehat{\mathsf{h}}(q)$ and $\widehat{\mathsf{c}}(q)$ are defined as $\widehat{h}_{ij}(q)=\rho\sqrt{x_i x_j}\widetilde{h}_{ij}(q)$ and $\widehat{c}_{ij}(q)=\rho\sqrt{x_i x_j}\widetilde{c}_{ij}(q)$, respectively.

As a generalization of Eq.\ \eqref{S(q)bis}, the static structure factor for the mixture, $S_{ij}(q)$, may be expressed in terms of
$\widetilde{h}_{ij}(q)$ as\cite{HM06}
\begin{equation}
\label{1.6}
S_{ij}(q)=x_i \delta_{ij}+\rho x_i x_j
\widetilde{h}_{ij}(q).
\end{equation}
In the particular case of a binary mixture ($\Nc=2$), rather than the
individual structure factors $S_{ij}(q)$, it is some combination of
them which may be easily associated with fluctuations of the
thermodynamic variables.\cite{AL67,BT70} Specifically, the
quantities\cite{HM06}
\begin{subequations}
\begin{equation}
\label{1.8}
S_{nn}(q)=S_{11}(q)+S_{22}(q)+2S_{12}(q),
\end{equation}
\begin{equation}
\label{1.9}
S_{nc}(q)=x_2 S_{11}(q)-x_1 S_{22}(q)+(x_2-x_1)S_{12}(q),
\end{equation}
\begin{equation}
\label{1.10}
S_{cc}(q)=x_2^2S_{11}(q)+x_1^2S_{22}(q)-2x_1
x_2S_{12}(q)
\end{equation}
\end{subequations}
are {associated with density fluctuations, density-concentration correlations, and concentration fluctuations, respectively}.

In terms of the structural properties, the energy route to the EOS of the mixture is given by
  \begin{equation}
  \label{energyMulti}
\frac{\langle E\rangle}{N}=\frac{3}{2}k_BT+\frac{\rho}{2}
\sum_{i,j=1}^{\Nc}x_i x_j\int d \mathbf{r}\, \phi_{ij}(r) g_{ij}(r) .
\end{equation}
In turn, the virial route is written as
\begin{equation}
\label{presMul}
Z=1-\frac{\betaT\rho}{6}
\sum_{i,j=1}^{\Nc}x_i x_j\int d \mathbf{r}\, r\frac{d \phi_{ij}(r)}{dr}g_{ij}(r) ,
\end{equation}
 while the compressibility route is given by
 \begin{equation}
\label{CompresMul}
\chi^{-1}=
\sum_{i,j=1}^{\Nc}\sqrt{x_i x_j}\left[\mathsf{I}+\widehat{\mathsf{h}}(0)\right]^{-1}_{ij}=1-\rho\sum_{i,j=1}^{\Nc} \widetilde{c}_{ij}(0).
\end{equation}

The $\mu$ route reads in this case\cite{S16,SR13}
\begin{subequations}
\begin{equation}
\betaT\mu_i=\ln\left(\rho x_i\Lambda_i^3\right)+\betaT\mu_i^\ex,
\end{equation}
\begin{equation}
\label{ChemPotMul}
\mu_i^\ex=\rho\sum_{j=1}^{\Nc}x_j \int_0^1 d \xi \int d \rr\, g_{\text{test}(i)j}^\xxx(r)
\frac{\partial \phi_{\text{test}(i)j}^\xxx(r)}{\partial\xi},
\end{equation}
\end{subequations}
where $\Lambda_i$ is the thermal de Broglie wavelength corresponding to particles of species $i$ and now $\xi$ is the coupling parameter of an extra test particle  of species $i$  to the rest  of the system, $\phi_{\text{test}(i)j}^\xxx(r)$ and $g_{\text{test}(i)j}^\xxx(r)$ being the potential and RDF, respectively, associated with the interaction between that test particle and a particle of species $j$.

Finally, the free-energy route for a mixture is\cite{S16}
\begin{subequations}
\begin{equation}
{\betaT a}=\sum_{i=1}^{\Nc}x_i\ln\left(\rho x_i\Lambda_i^3\right)-1+{\betaT a^\ex},
\end{equation}
\begin{equation}
\label{freeenergyMul}
{ a^\ex}=\frac{\rho}{2}\sum_{i,j=1}^{\Nc}x_i x_j \int_0^1 d \xi \int d \rr\,
g_{ij}^\xxx(r)\frac{\partial \phi_{ij}^\xxx(r)}{\partial\xi}.
\end{equation}
\end{subequations}
As in the one-component case, the common coupling parameter $\xi$ is now applied to all the pairs $(i,j)$.

\subsubsection{Approximate integral equation theories}

As said before, the OZ relation for a single-component fluid [Eq.\ \eqref{OZ}] cannot be closed with Eq.\ (\ref{Closureeq})  unless an approximation is introduced. Most of the approximations in liquid-state theory are made  by complementing  Eq.\ (\ref{Closureeq}) with  a \emph{closure} of the form $b(r)=\mathcal{B}[\gamma(r)]$, which yields $c(r)=\mathcal{C}[h(r)]$. Thus, one can obtain a \emph{closed integral equation} from the OZ relation, namely,
    \begin{equation}
    \label{IntEq}
  h(r)=\mathcal{C}[h(r)]+\rho \int d \mathbf{r}'\,\mathcal{C}[h(r')]h(|\rr-\rr'|).
  \end{equation}
It should be pointed out that, in contrast to a truncated density expansion, a closure is applied to \emph{all orders} in density. In any case, a given closure  consists of an \emph{ad hoc} approximation whose usefulness must be judged \emph{a posteriori}. The two prototype closures are the hypernetted-chain (HNC) closure\cite{M58,M60,vLGB59} and the Percus--Yevick (PY) closure.\cite{PY58} They are given by
\begin{equation}
\mathcal{B}[\gamma(r)]=\begin{cases}
0& (\text{HNC}),\\
\ln[1+\gamma(r)]-\gamma(r)& (\text{PY}),
\end{cases}
\end{equation}
or, equivalently,
\begin{equation}
\label{HNC-PY}
\mathcal{C}[h(r)]=\begin{cases}
h(r)-\ln [1+h(r)]-\betaT \phi(r)& (\text{HNC}),\\
[1+h(r)]\left[1-e^{\betaT\phi(r)}\right]& (\text{PY}).
\end{cases}
\end{equation}
From Eq.\ \eqref{IntEq}, these closures lead to the two prototype integral equations of liquid-state theory

As in the monocomponent case, the multicomponent OZ equation (\ref{OZMulti}) is not closed. The PY and HNC closures turn out to be straightforward generalizations of Eq.\ \eqref{HNC-PY}, namely,
\begin{equation}
c_{ij}(r)=\begin{cases}
h_{ij}(r)-\ln [1+h_{ij}(r)]-\betaT \phi_{ij}(r)& (\text{HNC}),\\
[1+h_{ij}(r)]\left[1-e^{\betaT\phi_{ij}(r)}\right]& (\text{PY}).
\end{cases}
\end{equation}

Two points should be stressed at this stage. On the one hand, the relationships between the thermodynamic and structural quantities quoted above remain strictly formal unless one is able to obtain explicit expressions of the RDF. On the other hand, due to the thermodynamic relationships between internal energy, pressure, isothermal susceptibility, and chemical potential [see Eq.\ \eqref{EpkappamufromF}], once the exact RDF  became available, the same result for the Helmholtz free energy $F$ should arise irrespective of the choice of the thermodynamic route used to obtain it, included the free-energy routes \eqref{freeenergy} and \eqref{freeenergyMul}. However, when an approximate RDF is used (e.g., the one obtained from the HNC or PY integral equations),  one gets (in general) a different approximate $F$ from each separate route (and for each separate $\xi$-protocol in the case of the $\mu$ and free-energy routes), thus leading to what is known as the
{\it thermodynamic consistency problem}.
In particular, it can be proved that the fourth virial coefficient, $B_4(T)$, obtained from the virial route in the HNC approximation is exactly equal to $\frac{3}{2}$ times the coefficient obtained from the compressibility route in the PY approximation,\cite{S16,SM10} regardless of the interaction potential, the number of components, and the dimensionality.

In this regard, it should be pointed out that other approximate closures to the OZ equations have been proposed\cite{V80,V81,MS83,RY84,BPGG86} in which an adjustable parameter is introduced to ensure thermodynamic consistency, usually between the virial and compressibility routes. Furthermore, it turns out that the energy and virial routes are equivalent in some approximate closures, such as the HNC approximation,\cite{BH76,M60} the linearized Debye--H\"uckel (LDH) approximation,\cite{S16,SFG09} and the Mean Spherical Approximation (MSA) for soft potentials.\cite{S16,S07a} The virial--energy equivalence in the cases of the LDH approximation and the MSA  for soft spheres extends to the free-energy route as well,\cite{S16} with independence of the choice of the charging protocol.

\subsection{HS fluids}

Let us now particularize the above results to the case of an \emph{additive} mixture of HS with an arbitrary number $\Nc$ of components.\cite{M08} In fact, our discussion will remain valid for $\Nc\to\infty$, {i.e.}, for polydisperse mixtures with a continuous distribution of sizes $\fx(\sigma)$. Of course if $\Nc=1$ one is considering the (monocomponent) HS fluid whose molecules have a diameter $\sigma$. For the HS mixture the intermolecular interaction potential reads
\begin{equation}
\label{HSpot}
\phi_{ij}(r)=\begin{cases}
\infty,&r<\sigma_{ij},\\
0,&r>\sigma_{ij},
\end{cases}
\end{equation}
in which the {additive}  hard core of the interaction between a sphere of species $i$ and a sphere of species $j$ is $\sigma_{ij}=\frac{1}{2}(\sigma _{i}+\sigma _{j})$, where the diameter of a sphere of species $i$ is $\sigma _{ii}=\sigma _{i}$. For later use, it is also convenient to introduce the packing fraction $\eta =\frac{\pi}{6}\rho \muM_3 $, where
\begin{equation}
\muM_n \equiv \langle \sigma^n\rangle=\sum_{i=1}^{\Nc} x_{i}\sigma
_{i}^{n}
\label{moments}
\end{equation}
 denotes the $n$th moment of the diameter distribution. Let us also introduce the \emph{reduced} moments $\mn\equiv \muM_n/\muM_1^n$.

\subsubsection{Thermodynamic routes}

Except for the compressibility route to the EOS [Eq.\ \eqref{CompresMul}], the different thermodynamic routes to the EOS simplify for the HS system. The energy and virial routes become
\begin{subequations}
\begin{equation}
\label{EnergyHS}
\langle E\rangle=\frac{3}{2}N k_BT,
\end{equation}
\begin{equation}
\label{ec1}
Z=1+
\frac{4 \eta}{\muM_3} \sum_{i,j=1}^{\Nc}x_i x_j\sigma_{ij}^3 \bar{g}_{ij},
\end{equation}
\end{subequations}
respectively, where the contact values of the RDF are defined as
\begin{equation}
\bar{g}_{ij}\equiv\lim_{r\to\sigma_{ij}^+}g_{ij}(r).
\end{equation}

As for the $\mu$ route, a natural choice for the interaction potential $\phi_{\text{test}(i)j}^\xxx(r)$ is a HS one characterized by a hard core $\sigma_{\text{test}(i)j}^\xxx$ with a linear dependence on the charging parameter $\xi$,
\begin{equation}
\sigma_{\text{test}(i)j}^\xxx=
\begin{cases}
  \xi\sigma_j,&0\leq\xi\leq\frac{1}{2},\\
  \left(\xi-\frac{1}{2}\right)\sigma_{i}+\frac{1}{2}\sigma_j,&\frac{1}{2}\leq \xi\leq 1.
\end{cases}
\end{equation}
Note that within the interval $0\leq\xi\leq\frac{1}{2}$, since $\sigma_{\text{test}(i)j}^\xxx\leq \frac{1}{2}\sigma_j$, the test particle can penetrate the other particles. The contribution associated with that interval can be evaluated exactly.\cite{SR13} On the other hand, in the interval $\frac{1}{2}\leq\xi\leq 1$ the test particle behaves with an additive diameter $\sigma_{\text{test}(i)}^\xxx=(2\xi-1)\sigma_i$. The final result is\cite{S16,SR13}
\begin{equation}
\label{ChemPotMulHS}
\betaT\mu_i^\ex=-\ln(1-\eta)+\frac{24\eta\sigma_i}{\muM_3}\sum_{j=1}^{\Nc}x_j \int_{\frac{1}{2}}^1 d \xi\,  \sigma_{\text{test}(i)j}^{\xxx^{\text{\scriptsize{2}}}}
\bar{g}_{\text{test}(i)j}^\xxx,
\end{equation}

In the case of the free-energy route [Eq.\ \eqref{freeenergyMul}], it seems natural to choose the intermediate potentials $\phi_{ij}^\xxx(r)$ as maintaining the HS structure but with a hard-core distance $\sigma_{ij}^\xxx$ interpolating between $\sigma_{ij}^{(\xi=0)}=0$ and $\sigma_{ij}^{(\xi=1)}=\sigma_{ij}$. Under those conditions one finds\cite{S16}
\begin{equation}
\label{freeenergyMulHS}
{\betaT a^\ex}=\frac{12\eta}{\muM_3}\sum_{i,j=1}^{\Nc}x_i x_j \int_0^1 d \xi\,
 \sigma_{ij}^{\xxx^{\text{\scriptsize{2}}}} \bar{g}_{ij}^\xxx\frac{\partial \sigma_{ij}^\xxx}{\partial\xi}.
\end{equation}
Here, the protocol $\sigma_{ij}^\xxx$ remains arbitrary. The simplest one is the distance rescaling $\sigma_{ij}^\xxx=\xi \sigma_{ij}$, a case in which  Eq.\ \eqref{freeenergyMulHS} becomes equivalent to the virial route [Eq.\ \eqref{ec1}].

It is clear from Eqs.\ \eqref{ec1}--\eqref{freeenergyMulHS} that the knowledge of the contact values  of the RDF  is sufficient to get the EOS of the mixture via the virial, $\mu$, and free-energy routes.
The expressions for the single-component fluid are of course obtained  when $\Nc=1$. For instance, the compressibility factor from the virial route \eqref{ec1} becomes
\begin{equation}
\label{virialHSmono}
Z=1+4 \eta \bar{g},\quad \bar{g}\equiv \lim_{r\to\sigma^+} g(r).
\end{equation}

From Eq.\ \eqref{EnergyHS} we note that the internal energy  of the multicomponent additive HS mixture is precisely that of an ideal-gas mixture. Since the internal energy per particle is independent of density, the energy route in the HS case is useless to derive the EOS. However, a physical meaning can be ascribed to the energy route if, first, it is applied
to a non-HS system that includes the HS one as a special case, and then, the HS limit is taken. When this process is undertaken via the square-shoulder potential, it turns out the the energy route becomes equivalent to the virial route.\cite{S16,S05,S06}

In connection with the virial expansion \eqref{pressure}, it should be pointed out that in the case of monocomponent HS fluids the virial coefficients are pure numbers independent of temperature, but for mixtures, they depend on composition. For the monocomponent system,  there exist exact values for the second, third, and fourth virial coefficients which date back to van der Waals and Boltzmann,\cite{NK72} while numerical values are available for $B_5$ to $B_{12}$, some of which are relatively  recent\cite{LKM05,CM06,W13} (see Table 3.9 of Ref.\ \onlinecite{S16}). In the case of mixtures, the second and third virial coefficients may be expressed  analytically in terms of the first three moments  as
\begin{subequations}
\label{Second&Thirdvir}
\begin{equation}
\label{Secondvir}
\bar{B}_2=1+3\frac{\mt}{\mth},
\end{equation}
\begin{equation}
\label{Thirdvir}
\bar{B}_3=1+6 \frac{\mt}{\mth}+3 \frac{\mt^3}{\mth^2},
\end{equation}
\end{subequations}
where $\bar{B}_n\equiv  B_n/(\frac{\pi}{6} M_3)^{n-1}$ is the reduced $n$th virial coefficient.
Unfortunately the fourth and higher virial coefficients for mixtures generally do not lend themselves to fully analytical expressions, although for some particular systems, such expressions for the fourth virial coefficient are available.\cite{S16} Hence, they have to be computed numerically and results for them are relatively scarce.

We should point out that the availability of only a few virial coefficients represents a restriction on the usefulness of the virial expansion and that many issues about it are still unresolved.\cite{M10} To begin with, even in the case that many more virial coefficients for HS systems were known, the truncated virial series for the corresponding compressibility factors would not be useful, in principle, for packing fractions higher than the one corresponding to the radius of convergence $\eta_\text{conv}$ of the whole series. Such a radius of convergence is determined by
the modulus of the singularity of $Z(\eta)$ closest to the origin in the complex plane, and this is not known, although lower bounds are available.\cite{LP64,FPS07} Further, although all  the available virial coefficients are positive, even the character of the series (either alternating or not) is still unknown. Results from higher dimensions suggest that the positive character might not be true for the higher virial coefficients of the HS fluid.\cite{CM06,RRHS08,AKV08}

\subsubsection{Cavity function for low densities}

It should be clear that obtaining the thermodynamic properties is, in general, simpler for HS systems than for other fluid systems. The same applies to the structural properties. In particular, the cavity function $y(r)$ for the single-component HS fluid is exactly known to second order in density,\cite{NvH52,RKM66,SM07,S16}
\begin{equation}
y(r)=1+y_1(r)\eta+y_2(r)\eta^2+\mathcal{O}(\eta^3),
\end{equation}
with
\begin{widetext}
\begin{subequations}
\begin{equation}
\label{y1r}
y_1(r)=\frac{1}{2}\left(2-{r^*}^2\right)\left(4+{r^*}\right)\Theta (2-{r^*}),
\end{equation}
\begin{align}
\label{y2r}
y_2(r)=&-\frac{36}{\pi}\left(\frac{3 {r^*}^6}{560}-\frac{{r^*}^4}{15}
+\frac{{r^*}^2}{2}-\frac{2{r^*}}{15}+\frac{9}{35{r^*}}\right)
\cos^{-1}\frac{-{r^*}^2+{r^*}+3}{\sqrt{3(4-{r^*}^2)}}\Theta(1-{r^*})
+\frac{2({r^*}-2)^2}{35r}\nonumber\\
&\times({r^*}^5+4{r^*}^4-51{r^*}^3-10{r^*}^2+479
{r^*}-81)\Theta(2-{r^*})
-\frac{({r^*}-3)^4}{35{r^*}}({r^*}^3+12{r^*}^2+27{r^*}-6)\Theta(3-{r^*})\nonumber\\
&+
\frac{18}{\pi}\left[-{r^*}^2\left(\frac{3{r^*}^2}{280}-\frac{41}{420}\right)\sqrt{3-{r^*}^2}
-\left(\frac{23}{15}{r^*}-\frac{36}{35{r^*}}\right)\cos^{-1}\frac{{r^*}}{\sqrt{3(4-{r^*}^2)}}
  \right.\nonumber\\
&  +\left(\frac{3
{r^*}^6}{560}-\frac{{r^*}^4}{15}+\frac{{r^*}^2}{2}+\frac{2{r^*}}{15}-\frac{9}{35{r^*}}\right)
\cos^{-1}\frac{{r^*}^2+{r^*}-3}{\sqrt{3(4-{r^*}^2)}}\nonumber\\ &
\left.+\left(\frac{3
{r^*}^6}{560}-\frac{{r^*}^4}{15}
+\frac{{r^*}^2}{2}-\frac{2{r^*}}{15}+\frac{9}{35{r^*}}\right)
\cos^{-1}\frac{-{r^*}^2+{r^*}+3}{\sqrt{3(4-{r^*}^2)}}\right]\Theta(\sqrt{3}-{r^*}).
\end{align}
\end{subequations}
\end{widetext}
Here, $\Theta(\cdot)$ is the Heaviside step function and  $r^*\equiv r/\sigma$. Thus, the contact value of the RDF is
\begin{align}
\bar{g}=&1+\frac{5\eta}{2}+
\left(\frac{2\,707}{280}+\frac{219\sqrt{2}}{140\pi}-\frac{4\,131\cos^{-1}\frac{1}{3}}{280\pi}\right)\eta^2\nonumber\\
&+\mathcal{O}(\eta^3).
\end{align}

\subsubsection{PY solution}
\label{subsubsec2B2}

\begin{table*}
\caption{Main thermodynamic quantities as obtained from the PY solution for monocomponent HS fluids via different routes, as well as from the CS approximation.
\label{tablePY}}
\begin{ruledtabular}
\begin{tabular}{lcccc}
Route&$ Z$&$\chi^{-1}$ &$\betaT\mu^\ex$&$\betaT a^\ex$\\
\hline
Virial
    &$\displaystyle{\frac{1+2\eta+3\eta^2}{(1-\eta)^2}}$
    &$\displaystyle{\frac{1+5\eta+9\eta^2-3\eta^3}{(1-\eta)^3}}$
    &$\displaystyle{\frac{2\eta(5-2\eta)}{(1-\eta)^2}}+2\ln(1-\eta)$
    &$\displaystyle{\frac{6\eta}{1-\eta}}+2\ln(1-\eta)$
    \\
Compressibility
    &$\displaystyle{\frac{1+\eta+\eta^2}{(1-\eta)^3}}$
    &$\displaystyle{\frac{(1+2\eta)^2}{(1-\eta)^4}}$
    &$\displaystyle{\frac{\eta(14-13\eta+5\eta^2)}{2(1-\eta)^3}-\ln(1-\eta)}$
    &$\displaystyle{\frac{3\eta(2-\eta)}{2(1-\eta)^2}-\ln(1-\eta)}$
    \\
$\mu$
    &$\displaystyle{-\frac{16-31\eta}{2(1-\eta)^2}}-\displaystyle{\frac{9}{\eta}\ln(1-\eta)}$
    &$\displaystyle{\frac{1+5\eta+9\eta^2}{(1-\eta)^3}}$
    &$\displaystyle{\frac{\eta(14+\eta)}{2(1-\eta)^2}-\ln(1-\eta)}$
    &$\displaystyle{\frac{3(6-\eta)}{2(1-\eta)}+\displaystyle{\frac{9-\eta}{\eta}}\ln(1-\eta)}$
    \\
CS
    &$\displaystyle{\frac{1+\eta+\eta^2-\eta^3}{(1-\eta)^3}}$
    &$\displaystyle{\frac{1+4\eta+4\eta^2-4\eta^3+\eta^4}{(1-\eta)^4}}$
     &$\displaystyle{\frac{\eta(8-9\eta+3\eta^2)}{(1-\eta)^3}}$
    &$\displaystyle{\frac{\eta(4-3\eta)}{\left(1-\eta\right)^2}}$
   \\
\end{tabular}
\end{ruledtabular}
\end{table*}

The PY integral equation has been solved exactly for both a monocomponent HS fluid\cite{T63,W63,W64,AL66} and a multicomponent additive HS mixture.\cite{L64,LZ71} For the former, such solution leads to the following valuable explicit results for structural properties:
\begin{subequations}
   \begin{equation}
    \label{gPYprime}
     \bar{g}=\frac{1+\eta/2}{(1-\eta)^2},\quad
\bar{g}'=-\frac{9}{2\sigma}\frac{\eta(1+\eta)}{(1-\eta)^3},
\end{equation}
    \begin{align}
    \frac{1}{S(q)}=&1+\frac{72\eta^2(2+\eta)^2}{(1-\eta)^4}{q^*}^{-4}+\frac{288\eta^2(1+2\eta)^2}{(1-\eta)^4}{q^*}^{-6}\nonumber\\
    &-\left[\frac{288\eta^2(1+2\eta)^2}{(1-\eta)^4}+\frac{72\eta^2(2-4\eta-7\eta^2)}{(1-\eta)^4}{q^*}^2\right.\nonumber\\
    &\left.+\frac{12\eta(2+\eta)}{(1-\eta)^2}{q^*}^4\right]\frac{\cos {q^*}}{{q^*}^{6}}-\left[\frac{288\eta^2(1+2\eta)^2}{(1-\eta)^4}\right.\nonumber\\
    &\left.-\frac{24\eta(1-5\eta-5\eta^2)}{(1-\eta)^3}{q^*}^{2}\right]  \frac{\sin {q^*}}{{q^*}^{5}},
    \label{10.22}
  \end{align}
   \begin{align}
     c(r)=&-\frac{\Theta(1-{r^*})}{(1-\eta)^4}\Big[{(1+2\eta)^2}-{6\eta(1+\eta/2)^2}{r^*}\nonumber\\
     &+  \frac{\eta(1+2\eta)^2}{2}{r^*}^3\Big].
   \label{10.23}
  \end{align}
\end{subequations}
In Eq.\ \eqref{gPYprime}, $\bar{g}'\equiv\lim_{r\to\sigma^+}\partial g(r)/\partial r$. Also, in Eq.\ \eqref{10.22}, $q^*\equiv q \sigma$.

Being an approximation, it is not surprising that the thermodynamic quantities predicted by the PY integral equation depend on the route followed. The main results are summarized in Table \ref{tablePY}. Note that, within a given route, the fundamental  thermodynamic relation
\begin{subequations}
\label{FET}
\begin{align}
\betaT\mu=&\frac{\partial(\eta\betaT a)}{\partial \eta}
\label{FET1}\\
=&{\betaT a}+Z
\label{FET2}
\end{align}
\end{subequations}
is satisfied.
We will come back to the thermodynamic consistency problem point later on. For the time being suffice it to mention now that a suitable combination of the virial ($Z_\text{PY,v}$) and the compressibility ($Z_\text{PY,c}$) EOS leads to a popular and rather accurate (in comparison with simulation results) EOS for the HS fluid, namely, the Carnahan--Starling (CS) EOS
\begin{align}
\label{ZCS}
Z_\text{CS}=&\frac{2}{3}Z_\text{PY,c}+\frac{1}{3}Z_\text{PY,v}\nonumber\\
=&\frac{1+\eta+\eta^2-\eta^3}{(1-\eta)^3},
\end{align}
which was derived in a totally independent manner by approximating the virial coefficients by integers and summing the resulting series.\cite{CS69} Other thermodynamic quantities stemming from Eq.\ \eqref{ZCS} are displayed in the last row of Table \ref{tablePY}. While there exist many other \emph{empirical} proposals for the EOS of the HS fluid,\cite{BN86,BC87,MGPC08,GB08,PBBCH19,TJM19} up to now the CS EOS stands out as perhaps the most successful \emph{simple} approximation.

\begin{figure}
\includegraphics[width=77mm]{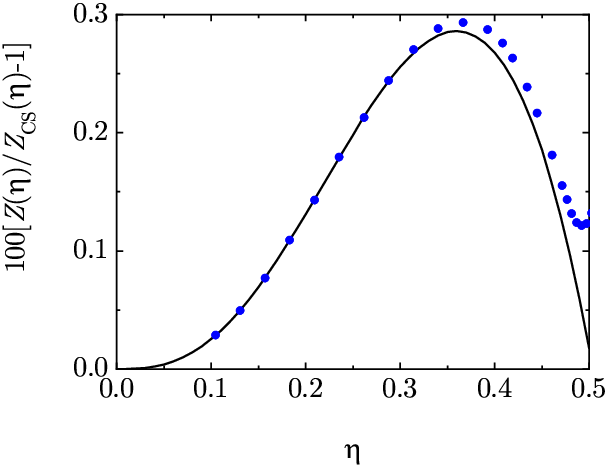}
\caption{
Plot of the relative deviation from the CS EOS, $100[Z(\eta)/Z_\text{CS}(\eta)-1]$. The solid line corresponds to the branch-point approximant \eqref{1bp}, while the circles represent MD simulation data from Ref.\ \protect\onlinecite{KLM04}. {Adapted from Fig.\ 2 of Ref.\ \onlinecite{SH09}}.\label{branch_point}}
\end{figure}

At this stage it is worthwhile recalling that the EOS for hard hyperspheres in odd dimensions greater than $3$ derived from the PY integral equation possesses a branch-point singularity on the negative real axis that is responsible for the radius of convergence and the alternating character of the virial series.\cite{FI81,L84,GGS90,BMC99,RS07,RRHS08,AKV08} It is very likely that these features are not artifacts of the PY approximation but would be shared by the exact EOS. However, in the case of the three-dimensional HS fluid, the radius of convergence of the PY EOS is artificially $\eta_{\text{conv}}=1$ and there is no
definite indication about the nature of the singularity responsible for the true radius of convergence or its value.\cite{CM06}
With this in mind,  a heuristic EOS for monocomponent HS systems was proposed\cite{SH09} that  relied  on the notion that the radius of convergence of the virial series might be dictated by a branch-point singularity. It reads
\begin{equation}
Z(\eta)=\frac{
1+\sum_{k=0}^3u_k\eta^k-u_0\left(1+2v_1\eta+v_2\eta^2\right)^{3/2}}{(1-\eta)^3},
\label{1bp}
\end{equation}
where $u_0$--$u_3$, $v_1$, and $v_2$   are parameters to be determined.
The functional form \eqref{1bp} is general enough as to include the PY EOS from the virial and compressibility routes, and thus also the CS EOS, by setting $v_2=v_1^2$, $u_1=1+3u_0 v_1$, $u_2=1+3u_0v_1^2$, and $u_3=b_4-19+u_0v_1^3$ with $b_4\equiv B_4/(\frac{\pi}{6}\sigma^3)^3=16$, $19$, and $18$ for $Z_{\text{PY,v}}$, $Z_{\text{PY,c}}$, and $Z_{\text{CS}}$, respectively. On the other hand, if the parameters in Eq.\ \eqref{1bp} are determined  by requiring agreement with the first seven virial coefficients, the resulting EOS\cite{SH09} successfully accounts for deviations of the CS EOS from molecular dynamics (MD) simulation values,\cite{KLM04} as shown in Fig.\ \ref{branch_point}.

 \begin{table*}
   \caption{Expressions for $z_2(\eta)$, $\betaT a_2(\eta)$, $X_2(\eta)$,  and $X_3(\eta)$, as obtained from the PY solution for multicomponent HS fluids via different routes, as well as from the BMCSL approximation.\label{tablePY2}}
\begin{ruledtabular}
\begin{tabular}{lcccc}
Route&$z_2(\eta)$&$\betaT a_2(\eta)$&$X_2(\eta)$&$X_3(\eta)$\\
\hline
Virial
    &$\displaystyle{\frac{3\eta^2}{(1-\eta)^2}}$&
    $\displaystyle{\frac{3\eta}{1-\eta}+3\ln(1-\eta)}$&
    $\displaystyle{\frac{9\eta}{1-\eta}+9\ln(1-\eta)}$&
    $\displaystyle{-\frac{3\eta(2-3\eta)}{(1-\eta)^2}-6\ln(1-\eta)}$\\
Compressibility
    &$\displaystyle{\frac{3\eta^2}{(1-\eta)^3}}$&
    $\displaystyle{\frac{3\eta^2}{2(1-\eta)^2}}$&
    $\displaystyle{\frac{9\eta^2}{2(1-\eta)^2}}$&
    $\displaystyle{\frac{3\eta^3}{(1-\eta)^3}}$\\
$\mu$
    &$\displaystyle{-\frac{9(2-{3}\eta)}{2(1-\eta)^2}-\frac{9}{\eta}\ln(1-\eta)}$&
    $\displaystyle{\frac{9(2-\eta)}{2(1-\eta)}+\frac{9}{\eta}\ln(1-\eta)}$&
    $\displaystyle{\frac{9\eta^2}{2(1-\eta)^2}}$&
    $0$\\
BMCSL
    &$\displaystyle{\frac{\eta^2(3-\eta)}{(1-\eta)^3}}$&
    $\displaystyle{\frac{\eta}{(1-\eta)^2}+\ln(1-\eta)}$&
    $\displaystyle{\frac{3\eta}{(1-\eta)^2}+3\ln(1-\eta)}$&
    $\displaystyle{-\frac{\eta(2-5\eta+\eta^2)}{(1-\eta)^3}-2\ln(1-\eta)}$\\
\end{tabular}
 \end{ruledtabular}
 \end{table*}

Now we turn to the multicomponent case.  The contact values of the RDF in the PY approximation are given by
\begin{equation}
\bar{g}_{ij}=\frac{1}{1-\eta
}+\frac{3}{2}\frac{\eta }{(1-\eta )^{2}}\frac{\sigma_i\sigma_j}{\sigma_{ij}}\frac{\muM_2}{\muM_3}.
\label{10.36}
\end{equation}

The thermodynamic quantities are of course sensitive to the route followed to derive them. On the other hand, regardless of the route, they have the following common form\cite{S16,HS16,HS18}
\begin{subequations}
  \begin{equation}
Z=\frac{1}{1-\eta}+\frac{3\eta}{(1-\eta)^2}\frac{\mt}{\mth}+z_2(\eta)\frac{\mt^3}{\mth^2},
\label{2.1}
\end{equation}
\begin{equation}
{\betaT a^\ex}=- \ln(1-\eta)+\frac{3\eta}{1-\eta}\frac{\mt}{\mth}+\betaT a_2(\eta)\frac{\mt^3}{\mth^2},
\label{aex}
\end{equation}
\begin{align}
\betaT \mu_i^\ex=&- \ln(1-\eta)+\frac{3\eta}{(1-\eta)^2}\frac{\mt}{\mth}\frac{\sigma_1}{\muM_1}\nonumber\\
&+\left[\frac{3\eta}{1-\eta}\frac{\mt}{\mth}+X_2(\eta)\frac{\mt^3}{\mth^2}\right]\frac{\sigma_i^2}{\muM_2}\nonumber\\
&+\left[\frac{\eta}{1-\eta}+\frac{3\eta^2}{(1-\eta)^2}\frac{\mt}{\mth}+X_3(\eta)\frac{\mt^3}{\mth^2}\right]\frac{\sigma_i^3}{\muM_3}.
\label{muiPY}
\end{align}
\end{subequations}
Therefore, only the coefficients of the combination of moments ${\mt^3}/{\mth^2}$ depend on the route.
The explicit expressions of $z_2$, $a_2$, $X_2$ and $X_3$, according to the PY virial, compressibility, and $\mu$  routes are displayed in Table \ref{tablePY2}.
Consistency with the single-component case (here denoted with the subscript $\pure$) gives
\begin{subequations}
\label{z2a2X2X3}
\begin{equation}
z_2(\eta)=Z_\pure(\eta)-\frac{1+2\eta}{(1-\eta)^2},
\end{equation}
\begin{equation}
\betaT a_2(\eta)={\betaT a^\ex_\pure}+\ln(1-\eta)-\frac{3\eta}{1-\eta},
\end{equation}
\begin{equation}
X_2(\eta)+X_3(\eta)={\betaT \mu^\ex_\pure}+\ln(1-\eta)-\frac{\eta(7-4\eta)}{(1-\eta)^2}.
\end{equation}
\end{subequations}
Taking into account Eq.\ \eqref{FET2}, the following relationship holds
\begin{equation}
X_2(\eta)+X_3(\eta)=\betaT a_2(\eta)+z_2(\eta).
\end{equation}
This in turn implies that the multicomponent extension of Eq.\ \eqref{FET} holds, namely,
\begin{subequations}
\label{FETMul}
\begin{align}
\sum_{i=1}^{\Nc}x_i\betaT\mu_i=&\frac{\partial(\eta\betaT a)}{\partial \eta}
\label{FETMul1}\\
=&{\betaT a}+Z.
\label{FETMul2}
\end{align}
\end{subequations}

\begin{figure}
\includegraphics[width=77mm]{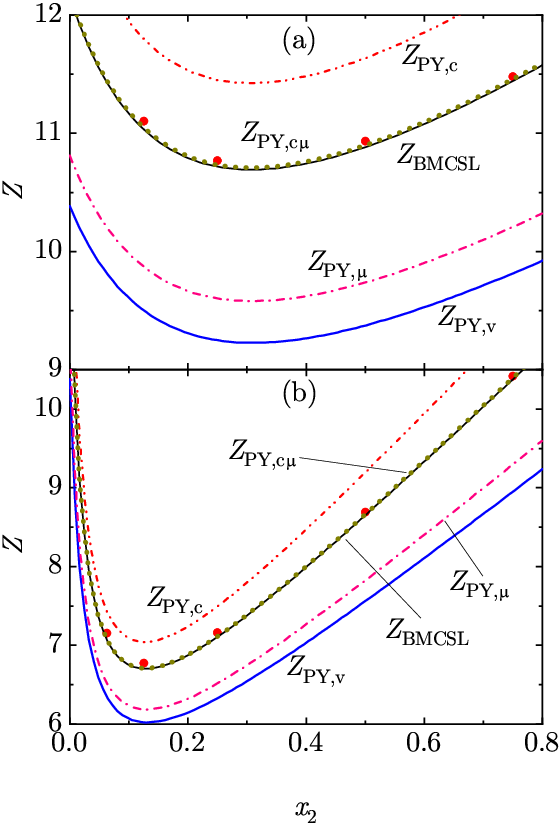}
\caption{
Plot of the compressibility factor $Z$ as a function of the mole fraction $x_2$ for a HS binary mixture with a packing fraction $\eta=0.49$ and  size ratios (a) $\sigma_1/\sigma_2=0.6$ and (b) $\sigma_1/\sigma_2=0.3$. Symbols are MC computer simulation values,\protect\cite{BMLS96} while the lines stand for (from top to bottom) $Z_{\text{PY,c}}$, $Z_{\text{PY,c}\mu}$ [see Eq.\ \protect\eqref{Zcmu}], $Z_{\text{BMCSL}}$, $Z_{\text{PY},\mu}$, and $Z_{\text{PY,v}}$, respectively. Note that $Z_{\text{PY,c}\mu}$ and $Z_{\text{BMCSL}}$ are hardly distinguishable. {Panels (a) and (b) are adapted from Figs.\ 7.9 and 7.10, respectively, of Ref.\ \onlinecite{S16}}.\label{Z_PY_mixt}}
\end{figure}

If an interpolation  between the virial and compressibility routes analogous to that of Eq.\ \eqref{ZCS} is carried out, one arrives at the  widely used and rather
accurate Boubl\'{\i}k--Mansoori--Carnahan--Starling--Leland (BMCSL)
EOS\cite{B70,MCSL71} for  HS mixtures. The associated coefficients $z_2$, $\betaT a_2$, $X_2$ and $X_3$ are also included in Table \ref{tablePY2}. They are consistent with Eq.\ \eqref{z2a2X2X3} if the single-component quantities are those of the CS EOS.
Since the $\mu$ route turns out to be more accurate than the virial route, it seems natural to propose an alternative interpolation formula as\cite{SR13,S16}
\begin{equation}
Z_{\text{PY,c}\mu}=\frac{11}{18}Z_{\text{PY,c}}+\frac{7}{18} Z_{\text{PY,v}}.
\label{Zcmu}
\end{equation}

As an assessment of  the performance of the compressibility factors related to the PY solution, Fig. \ref{Z_PY_mixt}  compares them  against Monte Carlo (MC) computer simulations\cite{BMLS96} for binary mixtures at $\eta=0.49$ and two values of the size ratio $\sigma_1/\sigma_2$.
It is observed that $Z_{\text{PY,v}}$  underestimates the simulation values, while $Z_{\text{PY,c}}$ overestimates them. The $\mu$ route compressibility factor, $Z_{\text{PY,c}\mu}$, lies below the simulation data, but, as said before, it exhibits a better  behavior than the virial route. The weighted average between $Z_{\text{PY,c}}$ and $Z_{\text{PY,v}}$ made in the construction of the BMCSL EOS  does a very good job. A slightly better agreement is obtained  from the weighted average between $Z_{\text{PY,c}}$ and $Z_{\text{PY,}\mu}$ [see Eq.\ \eqref{Zcmu}].

A comment is now in order. In the virial route, the starting point is the compressibility factor $Z$, as seen from Eq.\ \eqref{ec1}. Next, the Helmholtz free energy and chemical potential can be derived by standard thermodynamic relations, as summarized by the sequence
\begin{equation}
\label{fromZtomu}
\begin{array}{l}
Z\to\displaystyle{\int_0^1 dt\frac{Z(\eta t)-1}{t}}=\displaystyle{{\betaT a^\ex}}\\
\hspace{4cm}\downarrow\\
\hspace{2.5cm} \displaystyle{\mu_i^\ex=\frac{\partial  (\rho a^\ex)}{\partial \rho_i}}.
 \end{array}
\end{equation}
In the first step, use has been made of Eq.\ \eqref{Zfroma}.
In the case of the compressibility route [see Eq.\ \eqref{CompresMul}], use of the thermodynamic relation
\begin{equation}
\label{chi1}
\chi^{-1}=\frac{\partial \left(\eta Z\right)}{\partial \eta}
\end{equation}
allows one to obtain the compressibility factor  from the isothermal susceptibility as $ Z=\int_0^1 dt\,\chi^{-1}(\eta t)$. Thereafter, the sequence \eqref{fromZtomu} applies again.
On the other hand, in the $\mu$ route [Eq.\ \eqref{ChemPotMul}], one first obtains the chemical potential of any species $i$. Next, the free energy and the compressibility factor are found by means of the sequence
\begin{equation}
\label{frommutoZ}
\begin{array}{l}
\{\mu_i^\ex\}\to\displaystyle{\int_0^1dt\, \sum_{i=1}^{\Nc}x_i\mu_i^\ex(\eta t)}=\displaystyle{{a^\ex}}\\
\hspace{5cm}
\downarrow\\
\hspace{2cm} \displaystyle{Z-1=\sum_{i=1}^{\Nc}x_i\betaT \mu_i^\ex-{\betaT a^\ex}},
 \end{array}
\end{equation}
where in the first and second steps use has been made of Eqs.\ \eqref{FETMul1} and \eqref{FETMul2}, respectively.
Once the free energy is obtained from the $\mu$ route by the thermodynamic relation \eqref{FETMul1}, one might go back and derive the chemical potential of species $i$ via the thermodynamic relation in the second step of Eq.\ \eqref{fromZtomu}. As yet another instance of thermodynamic inconsistency, the resulting expression for $\betaT \mu_i^\ex$ differs from the original one in that $X_2\to X_2+\Delta X$ and $X_3\to X_3-\Delta X$ with $\Delta X=9(6-9\eta+2\eta^2)/2(1-\eta)^2+54\ln(1-\eta)/2\eta$.\cite{SR13}

\subsubsection{Phase behavior}
\label{sec2B4}

The investigation of the phase diagram of the HS fluid  has mainly relied on numerical simulations. Since the system is athermal,   the only parameter controlling the phase behavior of the single-component system is the density or, equivalently, the packing fraction. Presently, the accepted view is that the phase diagram of the HS system is constituted by four main branches. The first one, which goes from $\eta=0$ to the freezing packing fraction $\eta_\text{f}\simeq 0.492$,\cite{AW57,FMSV12} represents the stable fluid branch. There is then a tie line that joins the freezing point and the melting point at the packing fraction $\eta_\text{m} \simeq 0.543$\cite{HR68,FMSV12} in which there is fluid--solid coexistence. Above the melting point, the stable HS system is in a crystalline phase that ends at a close-packed face-centered cubic phase with a packing fraction $\eta_\text{cp}=\frac{\pi}{3 \sqrt{2}}$.\cite{S98b} There is also a \emph{metastable} fluid branch that extends past the freezing point and is conjectured to end at the densest possible random packing,\cite{T18} namely, with a ``jamming'' packing fraction   $\eta_\text{J}\simeq 0.64$. Finally, on the basis of experimental results  on colloidal HS\cite{vMU93,vBW95} (which easily form glasses) and some theoretical developments,\cite{S94b,Y95}  it has been concluded that a glass transition also occurs in the system at a packing fraction $\eta_\text{g}\simeq 0.58$ intermediate between $\eta_\text{f}$ and $\eta_\text{J}$---despite some controversy.

Evidence coming from the use of approximate EOS indicates that the freezing transition observed in computer simulations does not show up {as a singularity} in those approximations.\cite{AFLR84} Moreover, while it is quite plausible that  $Z(\eta)$ presents a singularity at the freezing density $\eta_\text{f}$,\cite{GJ80,J88,BET80} the
virial coefficients (or even their asymptotic behavior) do not seem to yield either any information concerning the freezing transition at $\eta_\text{f}$.\cite{GJ80} This might be related to the fact that $Z(\eta)$ remains finite when $\eta$ approaches $\eta_\text{f}$ from below.

Next, we turn to the high-density behavior. It should be remarked that the compressibility factor of the HS fluid, both for the stable and metastable fluid phases, is a monotonically increasing function of the packing fraction.\cite{M10} On the other hand, the fluid EOS, continued and extrapolated beyond the fluid--solid transition, is expected to have a divergence to infinity at a certain  packing fraction  $\eta=\eta_\infty$, i.e,
\begin{equation}
\lim_{\eta\to\eta_{\infty}}Z_{\text{fluid}}(\eta)=\infty,\quad
\eta_{\text{conv}}\leq\eta_\infty\leq \eta_{\text{cp}},
\label{17b}
\end{equation}
where we recall that $\eta_{\text{conv}}$ is the packing fraction corresponding to the radius of convergence of the virial series.
By studying the singularities of {Pad\'e approximants constructed from} the virial series for HS, Sanchez\cite{S94} came to the conclusion that such a singularity {was} related to the crystalline close-packing in these systems, i.e.,
\begin{equation}
\eta_\infty= \eta_{\text{cp}}.
\label{18}
\end{equation}
Other
authors\cite{W76,A75,A76,BL79,DS82,AFLPRR83,HvD84,GW88,WML91,SHY95,HSY98,WKV96,KV97,NAM00,GV01,W02,PV05,MMD06}
have also conjectured Eq.\ \eqref{18}.
{It is interesting to note that this conjecture was already suggested by Korteweg\cite{K92} and Boltzmann\cite{B98b} in the late 1800s.} On the other hand, the conjecture \eqref{18} has not been free from criticism\cite{GJ80}  {and some authors\cite{F72,AFLP82,MA86,SSM88,H97b,HY00} have
conjectured that $\eta_\infty=\eta_{\text{J}}$.\cite{B83} For a
{thorough} account of proposed EOS, including those enforcing $\eta_\infty= \eta_{\text{cp}}$ or $\eta_\infty=\eta_{\text{J}}$, see Ref.\  \onlinecite{MGPC08}.}
It has also been shown\cite{MSRH11} that the use of the direct Pad\'e approximants of the compressibility factor $Z$ is not reliable for the purpose of determining $\eta_\infty$. Instead, one should consider an approach in which the independent variable is the pressure rather than the density. The analysis shows that the knowledge of the first twelve virial
coefficients is not enough to decide whether $\eta_\infty=\eta_{\text{cp}}$ or $\eta_\infty=\eta_{\text{J}}$.

Once we have dealt with the one-component system, we now close this section by discussing the problem of fluid--fluid demixing in HS mixtures in which the knowledge about the virial coefficients is also valuable. An analysis of the solution of the PY integral equation for binary additive HS  mixtures\cite{LR64} leads to the conclusion
that no phase separation into two fluid phases exists in these
systems. The same conclusion is reached if one considers the   BMCSL EOS.\cite{B70,MCSL71} For a long time the belief was
that this was a true physical feature. Nevertheless, this belief
started to be seriously questioned after Biben and Hansen\cite{BH91} obtained fluid--fluid segregation in such mixtures
out of the solution of the OZ equation with the
Rogers--Young closure,\cite{RY84} {provided the size disparity was large enough}. More recently, an accurate EOS derived by invoking some consistency conditions\cite{S12c} does predict phase separation.
The importance of this issue resides in the
fact that if fluid--fluid phase separation occurs in additive HS
binary mixtures, it must certainly be entropy driven. In contrast,
in other mixtures such as molecular mixtures, {temperature plays a non-neutral role and demixing is a free-energy driven phase transition.}

The demixing problem has received a lot of attention in the
literature in different contexts and using different approaches.
For instance, Coussaert and Baus\cite{CB97,CB98} have proposed an EOS with improved virial behavior for a binary HS mixture that predicts a fluid--fluid transition at very
high pressures (metastable with respect to a fluid--solid one). On
the other hand, Regnaut {et al.}\cite{RDA01} have examined
the connection between empirical expressions for the contact
values of the pair distribution functions and the existence of
fluid--fluid separation in HS mixtures. Finally, in the
case of highly asymmetric binary additive HS mixtures, the
depletion effect has been invoked as the physical mechanism behind
demixing (see Refs.\ \onlinecite{DRE98,DRE99a,DRE99b,AA06,AWRE11}
and the bibliography therein) and an effective one-component fluid description has been employed.

In two instances, namely, the limiting cases of a pure HS
system  and that of a binary mixture in which one of the species
 consists of point particles, it is known that there is no fluid--fluid separation.\cite{V98} For size ratios $\sigma_1/\sigma_2$ other than $1$ or $0$, one can find the spinodal instability curve (whose minimum determines the  critical consolute point) by considering the truncated virial series. The results\cite{VM03,HT04,HML10} show that the values of
the critical pressure and packing fraction monotonically increase with
the truncation order. Extrapolation  of these values to infinite truncation order
suggests that the critical pressure diverges to infinity and the
critical packing fraction tends towards its close-packing value, thus
supporting a non-demixing scenario.  This
shows the extreme sensitivity of the demixing phenomenon to slight
changes in the approximate EOS that is chosen to describe the mixture.

The argument that the truncated virial series are prone to exhibit
demixing, albeit with larger and larger critical pressures, can be
reinforced by analyzing a binary mixture in which one of the species
 consists of point particles. In that limit, the virial coefficients of the mixture are
directly related to the ones of the pure fluid, which are known up to the twelfth.\cite{vR93, LKM05,CM05,CM06,W13} As mentioned previously, this system is known to
lack a demixing transition\cite{V98} but the truncated
virial series exhibits artificial critical points with the same
qualitative features as observed for the mixtures with nonzero size ratios.\cite{HML10}

Therefore, one can conclude that a \emph{stable} demixing fluid--fluid transition  {does not} occur in (three-dimensional) additive binary HS mixtures but it is preempted by a fluid--solid transition.\cite{DRE99b}

\section{The Rational Function Approximation (RFA) Method for the Structure of HS Fluids}
\label{sec3}

We have already pointed out that, apart from requiring, in general, hard numerical labor, a disappointing aspect of the usual approach to obtain $g(r)$, namely, the integral equation approach, in which the OZ equation is complemented by a closure relation between $c(r)$ and $h(r)$,\cite{BH76} is that the substitution of the (necessarily) approximate values of $g(r)$ obtained from them in the (exact) statistical--mechanical formulae may lead to  the thermodynamic consistency problem. In this section (which follows very closely material in Ref.\ \onlinecite{HYS08}), we describe the RFA method for HS fluids, which is an alternative
to the integral equation approach and in particular leads by
construction to thermodynamic consistency between the virial and compressibility routes.

\subsection{The single component HS fluid}
\label{ss_3.1}

\subsubsection{General framework}

We begin with the case of a HS single-component  fluid. The following presentation is equivalent to the
one given in Refs.\ \onlinecite{YS91,YHS96}, where all details can be
found, but more suitable than the former for direct generalization
to the case of mixtures.

The starting point will be the Laplace transform
\begin{equation}
\label{2.1G}
G(s)=\int_0^\infty  d r\,  e^{-sr}r g(r).
\end{equation}
The Fourier transform of the total correlation function is related to $G(s)$ by
\begin{align}
\label{new2}
\widetilde{h}(q)=&\frac{4\pi}{q}\int_0^\infty  d r \, r \sin(qr)
h(r)\nonumber\\
=&-2\pi \left.\frac{G(s)-G(-s)}{s}\right|_{s=\imath q}.
\end{align}

Without loss of generality, we can define an auxiliary function  $\Psi(s)$ through
\begin{equation}
\label{2.2}
G(s)=\frac{s}{2\pi}\left[\rho+ e^{s\sigma}\Psi(s)\right]^{-1}.
\end{equation}
The choice of $G(s)$ as the Laplace transform of $r g(r)$ and the
definition of $\Psi(s)$ from Eq.\ \eqref{2.2} are suggested by the
exact form of $g(r)$ [see Eq.\ \eqref{y1r}] to first order in density.\cite{YS91}

Since $g(r)=0$ for $r<\sigma$, while $\bar{g}=\text{finite}$, one
has
\begin{equation}
\label{3.2s}
g(r)=\Theta(r-\sigma)\left[\bar{g}+
(r-\sigma)\bar{g}'+\cdots\right].
\end{equation}
This property imposes a constraint
on the large-$s$ behavior of $G(s)$, namely,
\begin{equation}
\label{3.3s}
  e^{\sigma s}s G(s)=\sigma \bar{g}
+\left(\bar{g}+ \sigma\bar{g}'\right) s^{-1}+{\cal
O}(s^{-2}).
\end{equation}
Therefore, $\lim_{s\to\infty}  e^{s\sigma}sG(s)=\sigma
\bar{g}=\text{finite}$ or, equivalently,
\begin{equation}
\label{2.3}
\lim_{s\to\infty}s^{-2}\Psi(s)=\frac{1}{2\pi\sigma
\bar{g}}=\text{finite}.
\end{equation}
On the other hand,  according to Eq.\ \eqref{isocomp},
\begin{align}
\chi=&1-24\eta\sigma^{-3}\lim_{s\to 0}\frac{ \partial}{\partial
s}\int_0^\infty  d r\,  e^{-s r}r\left[g(r)-1\right]\nonumber\\
=&1-24\eta\sigma^{-3}\lim_{s\to 0}\frac{\partial}{\partial
s}\left[G(s)-s^{-2}\right].
\label{GG}
\end{align}
Since the isothermal susceptibility $\chi$ is also
finite, one has $\int_0^\infty  d
r\,r^2\left[g(r)-1\right]=\text{finite}$, so that the weaker
condition
 $\int_0^\infty  d
r\,r\left[g(r)-1\right]=\lim_{s\to 0}[G(s)-s^{-2}]=\text{finite}$
must hold. This in turn implies
\begin{equation}
\label{2.4}
e^{\sigma s}\Psi(s)=-\rho+\frac{s^3}{2\pi}+{\cal O}(s^5).
\end{equation}

Note that Eq.\ \eqref{2.2} can be formally rewritten as
\begin{equation}
\label{GGs}
G(s)=\frac{s}{2\pi}\sum_{n=1}^\infty
(-\rho)^{n-1}\left[\Psi(s)\right]^{-n}  e^{-ns\sigma}.
\end{equation}
Thus, the RDF is then given by
\begin{equation}
g\left({r}\right) =\frac{1}{2\pi r} \sum_{n=1}^{\infty
}(-\rho)^{n-1}\psi _{n}\left(r-n\sigma\right) \Theta
\left(r-n\sigma\right) ,
\label{g(r)}
\end{equation}
where
\begin{equation}
\psi_{n}\left( r\right) =\mathcal{L}^{-1}\left\{ s\left[\Psi
\left( s\right) \right] ^{-n}\right\} ,
\label{varphi}
\end{equation}
$\mathcal{L}^{-1}$ denoting the inverse Laplace transform.

Let us now derive an interesting property. For large $s$, Eq.\ \eqref{2.3} implies that $s\left[\Psi(s)\right]^{-n}\approx \left(2\pi\sigma \bar{g}\right)^n s^{-2n+1}$. This in turn implies the small-$r$ behavior $\psi_n(r)\approx \left(2\pi\sigma \bar{g}\right)^n r^{2n-2}/(2n-2)!$. Next, from Eq.\ \eqref{g(r)} we find that  the RDF exhibits $(2n-2)$th-order jump discontinuities at $r=n\sigma$, with $n=1,2,\ldots$, namely,
\begin{align}
\label{new8}
\lim_{r\to n\sigma^+}\frac{\partial^{2n-2}g(r)}{\partial r^{2n-2}}-&\lim_{r\to n\sigma^-}\frac{\partial^{2n-2}g(r)}{\partial r^{2n-2}}\nonumber\\
&=\frac{(-12\eta)^{n-1}\bar{g}^n}{\sigma^{2n-2}n}.
\end{align}
As exemplified by Eq.\ \eqref{y2r}, the exact RDF is also singular at some noninteger distances (in units of $\sigma$), such as $r=\sqrt{3}\sigma$.

Thus far, Eqs.\ \eqref{2.1G}--\eqref{new8} are formally exact, the auxiliary function $\Psi(s)$ remaining  unknown.

Let us now consider a \emph{class} of approximations where $\Psi(s)$ is an \emph{algebraic} function, so that  all the real-space functions $\psi_{n}\left( r\right)$ are \emph{regular} for $r>0$.
Under that condition, it is proved in Appendix \ref{appA}
that the the only singularities of $h(r)$ are located at $r=n\sigma$ (with $n=1, 2, \ldots$) and the DCF $c(r)$ is regular for any distance $r>\sigma$.

Among the class of algebraic-function approximations for $\Psi(s)$, let us focus on the subclass made of \emph{rational-function} approximations (RFA), namely,
\begin{equation}
\label{RFAn}
\Psi(s)=\frac{\sum_{\ell=0}^{\nu+2}\SE^{(\ell)} s^\ell}{\sum_{\ell=0}^{\nu}L^{(\ell)} s^\ell}.
\end{equation}
The difference between the degree of the numerator ($\nu+2$) and that of the denominator ($\nu$) is fixed by the physical requirement \eqref{2.3}. Since one of the coefficients in Eq.\ \eqref{RFAn} can be freely chosen, the number of independent coefficients is $2\nu+3$.
Next, the basic condition \eqref{2.4} fixes the first five coefficients in the expansion of $\Psi(s)$ in powers of $s$, so that $2\nu+3\geq 5$ or, equivalently, $\nu\geq 1$.

Enforcement of Eq.\ \eqref{2.4} up to order $s^2$ allows one to express $\SE^\zero$, $\SE^\one$, and $\SE^\two$ in terms of $L^\zero$, $L^\one$, and $L^\two$,
\begin{subequations}
\label{SEL}
\begin{equation}
\SE^\zero=-\rho L^\zero,
\end{equation}
\begin{equation}
\SE^\one=-\rho\left[L^\one-\sigma L^\zero\right],
\end{equation}
\begin{equation}
\SE^\two=-\rho\left[L^\two-\sigma
L^\one+\frac{1}{2} \sigma^2 L^\zero\right].
\end{equation}
\end{subequations}
As a consequence, insertion of \eqref{RFAn} into Eq.\ \eqref{2.2} yields
\begin{align}
\label{2.10j}
G(s)=&\frac{ e^{-\sigma s}}{2\pi s^2} \sum_{\ell=0}^{\nu}L^{(\ell)} s^\ell\nonumber\\
&\times \left[
\sum_{\ell=3}^{\nu+2}\SE^{(\ell)} s^{\ell-3}-\rho\sum_{\ell=0}^\nu\varphi_{2-\ell}(\sigma s)\sigma^{3-\ell} L^{(\ell)}\right]^{-1},
\end{align}
where
\begin{equation}
\label{2.9}
\varphi_\ell(x)\equiv \begin{cases}
\displaystyle{\frac{\sum_{k=0}^\ell \frac{(-x)^k}{k!}-
 e^{-x}}{x^{\ell+1}}},&\ell\geq 0,\\
 \displaystyle{-\frac{
 e^{-x}}{x^{\ell+1}}}, &\ell\leq -1
\end{cases}.
\end{equation}
Note that $\lim_{x\to 0}\varphi_\ell(x)=(-1)^\ell/(\ell+1)!$ and $0$ for $\ell\geq -1$ and $\ell\leq -2$, respectively.
The requirement \eqref{2.4} to orders $s^3$ and $s^4$ provide $L^\zero$ and $L^\one$ in terms of $L^\two$, $L^\three$, $L^\four$, $\SE^\three$, and $S^\four$. Therefore, $L^{(j)}$ and $\SE^{(j+1)}$ for $j\geq 2$ remain free.

By application of the residue theorem, the functions defined in Eq.\ \eqref{varphi} are explicitly given by
\begin{equation}
\psi _{n}\left( r\right) =
\sum_{m=1}^{n} \frac{\sum_{i=1}^{\nu+2}  \bar{\psi}_{mn}^{(i)}e^{s_{i}r}}{\left(n - m\right)!(m-1)! }
r^{n-m} ,
\label{z8bis}
\end{equation}
where
\begin{equation}
\bar{\psi}_{mn}^{(i)} \equiv \lim_{s \rightarrow s_{i}} \left(\frac{ \partial}{ \partial
s}\right)^{m-1} s\left[\frac{\Psi \left( s\right)}{s-s_i} \right] ^{-n},
\label{z8bisa}
\end{equation}
$s_{i}$ ($i=1,\ldots,\nu+2$) being the
roots of $\sum_{\ell=0}^{\nu+2}\SE^{(\ell)} s^\ell=0$.

\subsubsection{First-order approximation (PY solution)}

As seen above, the RFA \eqref{RFAn}
with the least number of coefficients to be determined corresponds to $\nu=1$, namely,
\begin{equation}
\label{2.5}
\Psi(s)=\frac{\SE^\zero+\SE^\one s+\SE^\two s^2+
s^3}{L^\zero+L^\one s},
\end{equation}
where we have chosen $\SE^\three=1$.   With such a choice and in view of
Eq.\ (\ref{2.4}), one finds Eq.\ \eqref{SEL} with $L^\two=0$  and
\begin{subequations}
\begin{equation}
\label{2.6}
L^\zero=2\pi \frac{1+2\eta}{(1-\eta)^2},
\end{equation}
\begin{equation}
\label{2.7}
L^\one=2\pi \sigma\frac{1+\eta/2}{(1-\eta)^2}.
\end{equation}
\end{subequations}
Finally, Eq.\ \eqref{2.10j} becomes
\begin{equation}
\label{2.8}
G(s)=\frac{ e^{-\sigma s}}{2\pi s^2} \frac{L^\zero+L^\one s}{
1-\rho\sum_{\ell=0}^1\varphi_{2-\ell}(\sigma s)\sigma^{3-\ell} L^{(\ell)}}.
\end{equation}

It is remarkable that  Eq.\ (\ref{2.8}), which has been derived here
as the simplest rational form for $\Psi(s)$ [see Eq.\ \eqref{RFAn}] complying with the
requirements \eqref{2.3} and \eqref{2.4}, coincides with the
solution to the PY closure of the OZ
equation\cite{W63} summarized in Sec.\ \ref{subsubsec2B2}.
It is clear that this first-order approximation is \emph{not} thermodynamically consistent.

\subsubsection{Second-order approximation}

\label{sec3A3}

In the  spirit of the RFA \eqref{RFAn}, the second simplest implementation corresponds to $\nu=2$, thus involving two new terms, namely,
\begin{equation}
\label{2.5bis}
\Psi(s)=\frac{\SE^\zero+\SE^\one s+\SE^\two s^2+
s^3+\alpha s^4}{L^\zero+L^\one s+L^\two s^2},
\end{equation}
where again we have chosen $\SE^\three=1$ and have called $\SE^\four=\alpha$.
Applying Eq.\ \eqref{2.4}, one finds Eq.\ \eqref{SEL} and
\begin{subequations}
\begin{equation}
\label{2.11}
L^\zero=2\pi \frac{1+2\eta}{(1-\eta)^2}
+\frac{12\eta}{1-\eta}\left[
\frac{\pi}{1-\eta}\frac{\alpha}{\sigma}-\frac{L^\two}{\sigma^2}\right],
\end{equation}
\begin{equation}
\label{2.12}
L^\one=2\pi
\sigma\frac{1+\frac{1}{2}\eta}{(1-\eta)^2} +\frac{2}{1-\eta}\left[
\pi\frac{1+2\eta}{1-\eta}\alpha-3\eta\frac{L^\two}{\sigma}\right].
\end{equation}
\end{subequations}
Furthermore,
\begin{equation}
\label{2.10}
G(s)=\frac{ e^{-\sigma s}}{2\pi s^2} \frac{L^\zero+L^\one s+{L^\two} s^2}{
1+\alpha s-\rho\sum_{\ell=0}^2\varphi_{2-\ell}(\sigma s)\sigma^{3-\ell} L^{(\ell)}}.
\end{equation}

Thus far, irrespective of the values of the coefficients $L^\two$
and $\alpha$, the conditions $\lim_{s\to\infty}
 e^{\sigma s}sG(s)=\text{finite}$ [see Eq.\ \eqref{3.3s}] and $\lim_{s\to
0}[G(s)-s^{-2}]=\text{finite}$ are satisfied. Of course, if
$L^\two=\alpha=0$, one recovers the PY approximation. More
generally, we may determine these two coefficients by prescribing the
compressibility factor $Z$ [or, equivalently, the contact value
$\bar{g}$, see Eq.\ \eqref{virialHSmono}] and  the isothermal susceptibility $\chi$. This gives
\begin{subequations}
\begin{equation}
\label{2.13}
{L^\two} ={2\pi \alpha \sigma}\bar{g},
\end{equation}
\begin{equation}
\label{2.14}
\chi=\left[\frac{2\pi}{L^\zero}\right]^2
\left[1-\frac{12\eta}{1-\eta}\frac{\alpha}{\sigma}\left(1+
\frac{2\alpha}{\sigma}\right)+\frac{12\eta}{\pi}\frac{\alpha
L^\two}{\sigma^3} \right].
\end{equation}
\end{subequations}
In order to ensure thermodynamic consistency, it is convenient to fix the isothermal susceptibility $\chi$ from $Z$ by means
of the thermodynamic relationship \eqref{chi1}.
Henceforth, we will restrict the use of the term RFA to this second-order approximation ($\nu=2$).

Upon substitution of Eqs.\
(\ref{2.11}) and (\ref{2.13}) into Eq.\ (\ref{2.14}) a quadratic
algebraic equation for $\alpha$ is obtained. The physical root is
\begin{equation}
\alpha=\frac{(1+2\eta)R}{(1-\eta)(3Z-1)+3[(1-\eta)Z-1-\eta]R},
\label{Ralfa}
\end{equation}
where
\begin{equation}
R\equiv\sqrt{1+\frac{Z-\frac{1}{3}}{Z-Z_{\text{PY,v}}}\left(
\frac{\chi}{\chi_{\text{PY,c}}}-1\right)}-1.
\label{S4}
\end{equation}
Here, $Z_{\text{PY,v}}$ and $\chi_{\text{PY,c}}$ correspond to the PY expressions in the virial and compressibility routes, respectively (see Table \ref{tablePY}).
The other root of the quadratic equation must be discarded because it corresponds to a
negative value of $\alpha$, which, according to Eq.\ \eqref{2.13},
yields a negative value of $L^\two$. This would imply the existence
of a positive real value of $s$ at which $G(s)=0$,\cite{YS91,YHS96} which is not compatible with a positive definite
RDF.

A reasonable compressibility factor must be $Z>Z_{\text{PY,v}}$. Moreover, if the chosen EOS
yields a diverging pressure at $\eta_\infty<1$, there must necessarily exist a certain packing fraction $\eta_\text{g}$ above which $\chi<\chi_{\text{PY,c}}$. As one approaches the value $\eta_\text{g}$ from below, so that $\chi/\chi_{\text{PY,c}}\to 1^+$, Eqs.\ \eqref{Ralfa} and \eqref{S4} show that $R,\alpha\to 0$, both quantities becoming complex beyond $\eta=\eta_\text{g}$. This may be interpreted as an indication that, at the
packing fraction $\eta_\text{g}$, the system
ceases to be a fluid and a glass transition occurs.\cite{YHS96,RHSY98,RH03}

Expanding \eqref{2.10} in powers of $s$ and using  Eq.\ (\ref{3.3s})
one can obtain the derivatives of the RDF at $r=\sigma^+$.\cite{RH97} In particular, the first derivative is
\begin{equation}
\bar{g}'=\frac{1}{2\pi\alpha\sigma}\left[L^\one-
L^\two\left(\frac{1}{\alpha}+\frac{1}{\sigma}\right)\right],
\label{3.19s}
\end{equation}
which may have some use in connection with perturbation theory.\cite{LL73}

It is worthwhile pointing out that the structure implied by Eq.\
(\ref{2.10}) coincides in this single-component case with the
solution of the Generalized Mean Spherical Approximation (GMSA),\cite{W73} where the OZ relation is solved under the ansatz that
the DCF has a Yukawa form outside the core.

For given $Z$ and $\chi$, once $G(s)$ has been fully determined, the RDF in real space is given by Eqs.\ \eqref{g(r)} and \eqref{z8bis} with $\nu=2$. Explicit expressions of $g(r)$ up to the second
coordination shell $\sigma\leq r\leq 3\sigma$ can be found in Ref.\
\onlinecite{DLS06}.
Furthermore, the  static structure factor $S(q)$ [{cf}. Eq.\
\eqref{S(q)bis}] and the Fourier transform $\widetilde{h}(q)$
 may be related to $G(s)$ from Eq.\ \eqref{new2} (see also Appendix \ref{appA}).
 Therefore, the basic structural quantities of the single-component HS
fluid, namely, the RDF and the static structure factor, may be
analytically determined within the RFA method once the
compressibility factor $Z$ (or, equivalently, the contact value
$\bar{g}$) is specified.

\begin{figure}
\includegraphics[width=77mm]{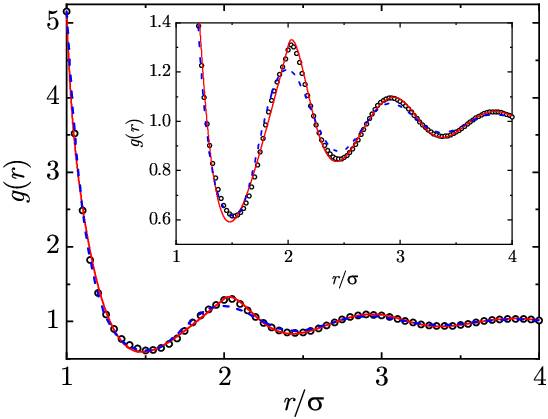}
\caption{RDF of a single-component HS fluid
for $\rho \sigma^3=0.9$ ($\eta=0.471$). The symbols represent MD simulation data,\protect\cite{KLM04} the dashed lines represent the results of the
approach of Ref.\ \protect\onlinecite{TNJH05}, and the solid lines
refer to those of the RFA method. The inset shows the oscillations
of $g(r)$ in more detail. {Adapted from Fig.\ 6.8 of Ref.\ \onlinecite{HYS08}}.}
\label{g(r)mono}
\end{figure}

In Fig.\ \ref{g(r)mono}, we compare MD
simulation data\cite{KLM04} of $g(r)$ for a density $\rho \sigma^3=0.9$ with the RFA prediction and the parameterized approach by
Trokhymchuk {et al.},\cite{TNJH05} where
$Z=Z_\text{CS}$ [{cf}.\ Eq.\ (\ref{ZCS})] and the
associated isothermal susceptibility $\chi=\chi_{\text{CS}}$ (see Table \ref{tablePY})
are taken in both cases. Both theories are rather accurate, but the
RFA captures better the maxima and minima of $g(r)$.\cite{HSY06}

It is also possible to obtain the DCF $c(r)$ within the RFA method. Using Eqs.\ \eqref{OZFourier} and
\eqref{new2} [see also Eq.\ \eqref{A9C}], and applying the residue theorem, one gets, after some
algebra,
\begin{align}
\label{c(r)}
c(r)=&\Bigg(\aK_+\frac{ e^{\kappa r^*}}{r^*}+\aK_-\frac{ e^{-\kappa
r^*}}{r^*}+\frac{\aK_{-1}}{r^*}+\aK_0+\aK_1 r^*\nonumber\\
&+\aK_3 {r^*}^3\Bigg)
\Theta(1-r^*)+\aK\frac{ e^{-\kappa r^*}}{r^*},
\end{align}
where
\begin{equation}
\kappa=\frac{\sigma}{\alpha}\sqrt{\frac{12\eta\alpha L^\two}{\pi\sigma^3}+1-\frac{12\alpha}{\sigma}\left(1+
\frac{2\alpha}{\sigma}\right)\frac{\eta}{1- \eta}}
\label{kappa}
\end{equation}
and the expressions for the amplitudes can be found in Appendix \ref{appB}.
In contrast to the PY result [Eq.\  \eqref{10.23}], now the DCF does not vanish outside the
hard core ($r > \sigma$) but has a Yukawa form in that region.
Note that Eq.\ \eqref{aprima} guarantees that
$c(0)=\text{finite}$, while Eq.\ \eqref{eca3} yields
$\lim_{r\to\sigma^+}c(r)-\lim_{r\to\sigma^-}c(r)=L^\two/2\pi\alpha=\bar{g}$. The latter proves the continuity of the  indirect correlation function
$\gamma(r)$ at $r=\sigma$. With the above results
[Eqs.\ \eqref{g(r)} and \eqref{c(r)}],  one may immediately write the
function $\gamma(r)$. Finally, we note that, according to Eq.\ \eqref{Closureeq}, the bridge function
$b(r)$ is linked to $\gamma(r)$ and $y(r)$  through
\begin{equation}
 b(r)= \ln y(r)-\gamma(r).
\label{b(r)}
\end{equation}
Thus, within the RFA method, the bridge function is also
completely specified analytically for $r\geq \sigma$ once $Z$ is
prescribed.

\begin{figure}
\includegraphics[width=77mm]{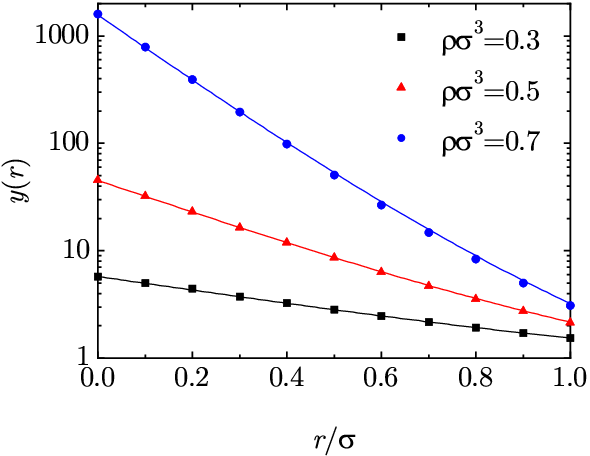}
\caption{Cavity function  of a single-component HS fluid in the
overlap region for $\rho \sigma^3=0.3$, $0.5$, and $0.7$. The solid
lines represent our proposal \protect\eqref{lny} with
$Z=Z_{\text{CS}}$, while the symbols represent MC simulation results.\protect\cite{LM84} {Adapted from Fig.\ 6.9 of Ref.\ \onlinecite{HYS08}}.}
\label{y(r)_hs}       
\end{figure}

\begin{figure}
\includegraphics[width=77mm]{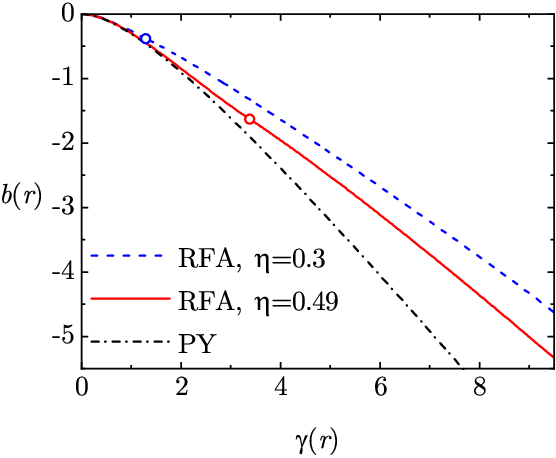}
\caption{Parametric plot of the bridge function $b(r)$ vs the
indirect correlation function $\gamma(r)$. The dashed line refers to
the RFA for $\eta=0.3$, while the solid line refers to the RFA for
$\eta=0.49$. In each case, the branch of the curve to the right of
the circle corresponds to $r\leq \sigma$, while that to the left
corresponds to $r\geq \sigma$. For comparison, the density-independent PY closure
$b(r)=\ln[1+\gamma(r)]-\gamma(r)$ is also plotted (dash-dotted
line). {Adapted from Fig.\ 6.10 of Ref.\ \onlinecite{HYS08}}.}
\label{B_vs_gamma}       
\end{figure}

If one wants to have $b(r)$ also for $0\leq r \leq \sigma$, then an
expression for the cavity function in that region is required. Here,
we propose such an expression using a limited number of constraints.
First, since the cavity function and its first derivative are
continuous at $r=\sigma$, we have
\begin{equation}
y(\sigma)=\bar{g},\quad
\frac{y'(\sigma)}{y(\sigma)}=\frac{L^\one}{L^\two}-\frac{1}{\alpha}-\frac{1}{\sigma},
\label{yp1}
\end{equation}
where Eqs.\ \eqref{2.13} and \eqref{3.19s} have been used. Next, we consider the following exact
zero-separation theorems:\cite{L95,LGL96,LM01}
\begin{subequations}
\label{zst}
\begin{equation}
\ln y(0)=Z-1+\int_0^1
dt\frac{Z(\eta t)-1}{t},
\label{36zst}
\end{equation}
\begin{equation}
\frac{y'(0)}{y(0)}=-6\eta \frac{y(\sigma)}{\sigma}.
\label{37zst}
\end{equation}
\end{subequations}
The four conditions \eqref{yp1}--\eqref{zst} can be enforced by
assuming a cubic polynomial form for $\ln y(r)$ inside the core,
namely,
\begin{equation}
y(r)=\exp\left(Y_0+Y_1 r^*+Y_2 {r^*}^2+Y_3 {r^*}^3\right), \quad 0\leq r\leq
\sigma,
\label{lny}
\end{equation}
where
\begin{subequations}
\begin{equation}
Y_0=Z-1+\int_0^1
dt\frac{Z(\eta t)-1}{t},
\end{equation}
\begin{equation}
Y_1=-6\eta y(\sigma),
\end{equation}
\begin{equation}
Y_2=3\ln y(\sigma)-\frac{\sigma y'(\sigma)}{y(\sigma)}-3Y_0-2Y_1,
\end{equation}
\begin{equation}
Y_3=-2\ln y(\sigma)+\frac{\sigma y'(\sigma)}{y(\sigma)}+2Y_0+Y_1.
\end{equation}
\end{subequations}
The proposal \eqref{lny} is compared with available MC data\cite{LM84} in Fig.\ \ref{y(r)_hs}, where an excellent agreement can
be observed.

Once the cavity function $y(r)$ provided by the RFA method is
complemented by Eq.\ \eqref{lny}, the bridge function $b(r)$ can be
obtained at any distance. Figure \ref{B_vs_gamma} presents a
parametric plot of the bridge function vs the indirect
correlation function as given by the RFA method for two different
packing fractions, as well as the result associated with the PY
closure. The fact that one gets a smooth curve means that, within the RFA, the oscillations in $\gamma(r)$ are highly correlated with those of $b(r)$. Further, the effective closure relation in the RFA turns out to be density dependent, in contrast with what occurs for the PY theory. Note that the absolute value $|b(r)|$  for a given value of $\gamma(r)$ is smaller in the RFA than the PY value. On the other hand, the RFA and PY curves become  closer as the density increases.
Since the PY theory is known to yield rather poor values of the
cavity function inside the core,\cite{MS06,SM07} it seems likely
that the present differences may represent yet another manifestation
of the superiority of the RFA method.

\subsection{The multicomponent HS fluid} \label{ss_3.2}
The method outlined in Subsection \ref{ss_3.1} will be now extended
to an $\Nc$-component mixture of additive HS.
Similar to what we did in the single-component case,  we introduce
the Laplace transforms of $r g_{ij}(r)$,
\begin{equation}
\label{3.1}
G_{ij}(s)=\int_0^\infty  d r\,  e^{-sr}r g_{ij}(r),
\end{equation}
so that the Fourier transform of $h_{ij}(r)$ can be obtained as
\begin{equation}
\widetilde{h}_{ij}(q)=-2\pi \left.\frac{G_{ij}(s)-G_{ij}(-s)}{s}
\right|_{s=\imath q}.
\label{1.7}
\end{equation}
The counterparts of Eqs.\ \eqref{3.2s} and \eqref{3.3s} are
\begin{subequations}
\begin{equation}
\label{3.2}
g_{ij}(r)=\Theta(r-\sigma_{ij})\left[\bar{g}_{ij}+
(r-\sigma_{ij})\bar{g}_{ij}'+\cdots\right],
\end{equation}
\begin{equation}
\label{3.3}
  e^{\sigma_{ij}s}s G_{ij}(s)=\sigma_{ij}\bar{g}_{ij}
+\left(\bar{g}_{ij}+\sigma_{ij}\bar{g}_{ij}'\right) s^{-1}+{\cal O}(s^{-2}).
\end{equation}
\end{subequations}
Moreover, according to Eq.\ \eqref{CompresMul}, the condition of a finite compressibility implies that
$\widetilde{h}_{ij}(0)=\text{finite}$. As a consequence, for small
$s$,
\begin{equation}
\label{3.4}
s^2
G_{ij}(s)=1+H_{ij}^{(0)}s^2+H_{ij}^{(1)}s^3+\cdots
\end{equation}
with $H_{ij}^\zero=\text{finite}$ and
$H_{ij}^\one=-\widetilde{h}_{ij}(0)/4\pi=\text{finite}$, where
\begin{equation}
\label{3.5}
H_{ij}^\n\equiv \frac{1}{n!}\int_0^\infty  d r\, (-r)^n r h_{ij}(r).
\end{equation}

 We are now in the position to generalize the approximation
(\ref{2.10}) to the $\Nc$-component case.\cite{YSH98,MYSH07,YSH08} While such a
generalization may be approached in a variety of ways,  two
motivations are apparent. On the one hand, we want to recover the PY
result\cite{L64} as a particular case in much the same fashion as
in the single-component system. On the other hand, we want to
maintain the development as simple as possible. Taking all of this
into account, we generalize Eq.\ \eqref{2.10} as
\begin{equation}
\label{3.6}
G_{ij}(s)=\frac{ e^{-\sigma_{ij} s}}{2\pi s^2}
\left({\sf L}(s)\cdot \left[(1+\alpha s)\mathsf{I}-{ \sf
A}(s)\right]^{-1}\right)_{ij},
\end{equation}
where  $\mathsf{L}(s)$ and $\mathsf{A}(s)$ are the matrices
\begin{equation}
\label{3.7}
L_{ij}(s)=L_{ij}^\zero+L_{ij}^\one
s+L_{ij}^\two s^2,
\end{equation}
\begin{equation}
\label{3.8}
A_{ij}(s)=\rho_i\sum_{\ell=0}^2\varphi_{2-\ell}(\sigma_i s)\sigma_i^{3-\ell} L_{ij}^{(\ell)},
\end{equation}
the functions $\varphi_\ell(x)$ being defined by Eq.\ \eqref{2.9}. We
note that, by construction, Eq.\ (\ref{3.6}) complies with the
requirement $\lim_{s\rightarrow\infty}  e^{\sigma_{ij}s}s
G_{ij}(s)=\text{finite}$. Further, in view of Eq.\ (\ref{3.4}), the
coefficients of $s^0$ and $s$ in the power series expansion of $s^2
G_{ij}(s)$ must be $1$ and $0$, respectively. This yields $2\Nc^2$
conditions that allow us to express ${\sf L}^\zero$ and ${\sf
L}^\one$ in terms of ${\sf L}^\two$ and $\alpha$. The solution is\cite{YSH98}
\begin{subequations}
\label{3.13&3.14}
\begin{equation}
\label{3.13}
L_{ij}^\zero=\lambdak+\lambdakk\sigma_j+2\lambdakk\alpha-
\lambdak\sum_{k=1}^{\Nc} \rho_k\sigma_k L_{kj}^\two,
\end{equation}
\begin{align}
\label{3.14}
L_{ij}^\one=&\lambdak\sigma_{ij}+\frac{1}{2}\lambdakk\sigma_i\sigma_j
+(\lambdak+\lambdakk\sigma_i)\alpha\nonumber\\
&-\frac{1}{2}\lambdak\sigma_i
\sum_{k=1}^{\Nc} \rho_k\sigma_k L_{kj}^\two,
\end{align}
\end{subequations}
where $\lambdak\equiv 2\pi/(1-\eta)$ and $\lambdakk\equiv
6\pi(\mt/\mth)\eta/(1-\eta)^2$.

In parallel with the development of the single-component case, ${\sf
L}^\two$ and $\alpha$ can be chosen arbitrarily. Again, the choice
$L_{ij}^\two=\alpha=0$ gives the PY solution.\cite{L64,BH77} Since
we want to go beyond this approximation, we will determine those
coefficients by taking prescribed values for $\bar{g}_{ij}$,
which in turn, via Eq.\ \eqref{ec1}, give the EOS of the mixture. This
also leads to the required value of $\chi$ via Eq.\ \eqref{chi1}, thus making the theory thermodynamically
consistent. In particular, according to Eq.\ (\ref{3.3}),
\begin{equation}
\label{3.17}
{L_{ij}^\two}={2\pi\alpha\sigma_{ij}} \bar{g}_{ij}.
\end{equation}
The condition related to $\chi$ is more involved. Making use of Eq.\
(\ref{3.4}), one can get $\widehat{h}_{ij}(0)=-4\pi \rho\sqrt{x_ix_j}H_{ij}^\one$
in terms of ${\sf L}^\two$ and $\alpha$, and then insert it into Eq.\
\eqref{CompresMul}. Finally, elimination of $L_{ij}^\two$ in favor of
$\alpha$ from Eq.\ (\ref{3.17}) produces a polynomial equation of
degree $2\Nc$, whose physical root is determined by the requirement
that $G_{ij}(s)$ is positive definite for  positive real $s$. It
turns out that the physical solution corresponds to the smallest of
the real roots. Once $\alpha$ is known, upon substitution into Eqs.\
(\ref{3.6}), (\ref{3.13&3.14}),  and (\ref{3.17}), the
scheme is complete. Also, using Eq.\ (\ref{3.3}), one can easily
derive the result
\begin{equation}
\label{3.19}
\bar{g}_{ij}'=\frac{1}{2\pi\alpha\sigma_{ij}}\left[L_{ij}^\one-
L_{ij}^\two\left(\frac{1}{\alpha}+\frac{1}{\sigma_{ij}}\right)\right].
\end{equation}
It is straightforward to check that the results of Sec.\ \ref{sec3A3} are recovered  by setting $\sigma_i=\sigma$, regardless
of the values of the mole fractions and the number of components.

Once $G_{ij}(s)$ has been determined, inverse Laplace transformation
directly yields $rg_{ij}(r)$. Although, in principle, as in the single-component case, this can be done
analytically, it is more practical to use one of the efficient
methods discussed by  Abate and Whitt\cite{AW92} to numerically
invert Laplace transforms.\footnote{A code using the Mathematica computer algebra system to obtain
$G_{ij}(s)$ and $g_{ij}(r)$ with the present method is available
from the web page
\href{https://www.eweb.unex.es/eweb/fisteor/santos/filesRFA.html}{https://www.eweb.unex.es/eweb/fisteor/santos/filesRFA.html}}

{}From Eqs.\ \eqref{OZMulti2} and \eqref{1.7} it is possible to obtain explicit expressions for $\widetilde{c}_{ij}(q)$.
Subsequent numerical inverse Fourier transformation yields $c_{ij}(r)$.
The indirect correlation functions $\gamma_{ij}(r) \equiv h_{ij}(r)
- c_{ij}(r)$ readily follow from the previous results for the RDF
and DCF. Finally, in this case the bridge
functions $b_{ij}(r)$ for $r>\sigma_{ij}$ are linked to $g_{ij}(r)$
and $c_{ij}(r)$ through\cite{YSH00a} $b_{ij}(r)= \ln g_{ij}(r)-\gamma_{ij}(r)$
 and so  we have a full set of  structural properties of a
 multicomponent fluid mixture of HS once the contact
 values $g_{ij}(\sigma_{ij})$ are specified.

 \begin{figure}
\includegraphics[width=77mm]{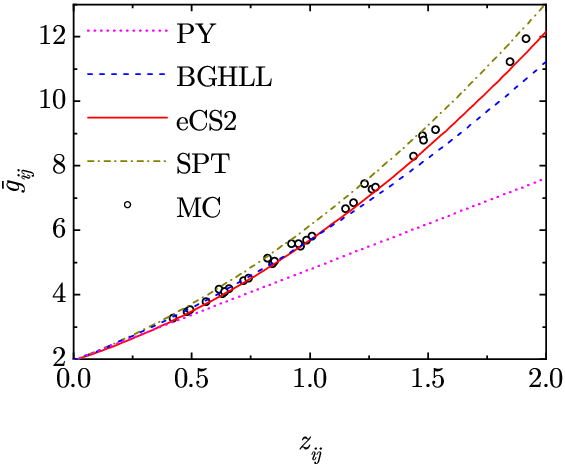}
\caption{
Plot of the contact value $\bar{g}_{i j}$ as a function of the
parameter $z_{i j}$ [see Eq.\ \eqref{zij}]. The circles are MC simulation
data\cite{MMYSH02} for ternary additive HS
mixtures with diameter ratios  $\sigma_2/\sigma_1=2$ and $\sigma_3/\sigma_1=3$ at a packing fraction $\eta=0.49$ and mole fractions  $(x_1,x_2,x_3)=(0.70,0.20,0.10)$, $(0.60,0.20,0.30)$, $(0.86,0.11,0.03)$, $(0.85,0.05,0.10)$, and $(0.90,0.07,0.03)$. Lines are theoretical predictions: from bottom  to top---PY [Eq.\ \eqref{10.36}], BGHLL [Eq.\ \eqref{15BGHLL}], eCS2 [Eq.\ \eqref{15eCS2}], and SPT [Eq.\ \eqref{15SPT}]. {Adapted from Fig.\ 3 of Ref.\ \onlinecite{MMYSH02}}.
\label{g_vs_z}}
\end{figure}

\begin{figure}
\includegraphics[width=75mm]{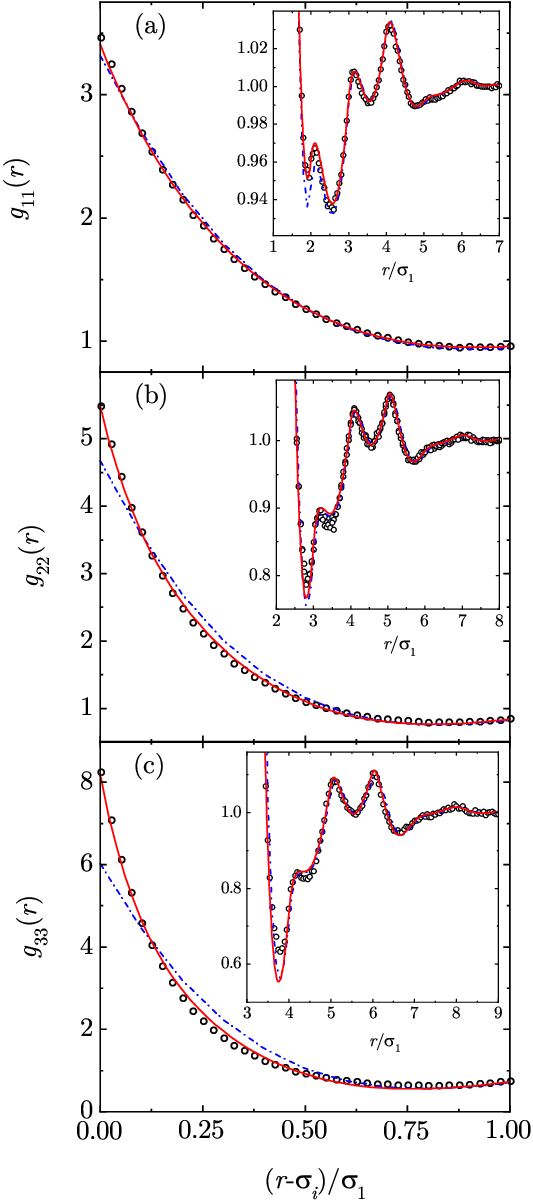}
\caption{
RDF (a) $g_{11}(r)$, (b) $g_{22}(r)$, and (c) $g_{33}(r)$ in the range $\sigma_i\leq r\leq\sigma_i+\sigma_1$ for a ternary HS
mixture with diameter ratios  $\sigma_2/\sigma_1=2$ and $\sigma_3/\sigma_1=3$
at a packing fraction $\eta=0.49$ and mole fractions $(x_1,x_2,x_3)=(0.7,0.2,0.1)$. The circles are MC simulation results,\protect\cite{MMYSH02} the solid lines are RFA predictions, and
the dash-dotted lines are PY predictions. The insets show in detail the oscillations in the region $\sigma_i\leq r\leq\sigma_i+6\sigma_1$.
{Adapted from Fig.\ 4 of Ref.\ \onlinecite{MMYSH02}}.
\label{ternary}}       
\end{figure}

The PY prediction given by Eq.\ \eqref{10.36} underestimates the contact values  and is not particularly accurate. On the other hand, the contact values obtained from  the scaled particle theory (SPT) for mixtures,\cite{RFL59,LHP65,MR75,R88,HC04} namely,
\begin{align}
\bar{g}_{ij}^{\text{SPT}}=&\frac{1}{1-\eta
}+\frac{3}{2}\frac{\eta }{(1-\eta )^{2}}\frac{\sigma_i\sigma_j}{\sigma_{ij}}
\frac{\muM_2}{\muM_3}
\nonumber\\&
+\frac{3}{4}\frac{\eta^{2}}{(1-\eta)^{3}}\left(\frac{\sigma_i\sigma_j}{\sigma_{ij}}
\frac{\muM_2}{\muM_3}\right)^{2},
\label{15SPT}
\end{align}
tend to overestimate the simulation values.
Interestingly, insertion of Eq.\ \eqref{15SPT} into Eq.\ \eqref{ec1} yields an EOS that coincides with that of the PY solution in the compressibility route.
Boubl\'{\i}k\cite{B70} and, independently, Grundke
and Henderson\cite{GH72} and Lee and Levesque\cite{LL73} proposed
an interpolation between the PY and the SPT contact values, that we
will refer to as the BGHLL values,
\begin{align}
\bar{g}_{ij}^{\text{BGHLL}}=&\frac{1}{1-\eta
}+\frac{3}{2}\frac{\eta }{(1-\eta )^{2}}\frac{\sigma_i\sigma_j}{\sigma_{ij}}
\frac{\muM_2}{\muM_3}\nonumber\\
&+\frac{1}{2}\frac{\eta^{2}}{(1-\eta)^{3}}\left(\frac{\sigma_i\sigma_j}{\sigma_{ij}}
\frac{\muM_2}{\muM_3}\right)^{2}.
\label{15BGHLL}
\end{align}
This leads, via Eq.\  \eqref{ec1}, to the widely used and rather
accurate BMCSL EOS.\cite{B70,MCSL71}

Refinements of the BGHLL values have been subsequently introduced, among others, by Henderson {et al.},\cite{HMLC96,YYH96,HC98,HC98b,HC00,HBCW98,MHC99,CCHW00,SYHAH09} Matyushov and
Ladanyi,\cite{ML97} and Barrio and Solana\cite{BS00} to eliminate
some drawbacks of the BMCSL EOS in the so-called colloidal limit of
{binary} HS mixtures. On a different path, but also having to do
with the colloidal limit, Viduna and Smith\cite{VS02a,VS02b} have proposed a method to obtain contact values of the RDF of HS mixtures from a given EOS. However, most of these proposals are not easily generalized to the case of an arbitrary number of components. Yet another extension of the CS contact value to a mixture with an arbitrary number of components, different from Eq.\ \eqref{15BGHLL}, was proposed in Ref.\ \onlinecite{SYH02} (see also Sec.\ \ref{sec:e2}), specifically
\begin{align}
\bar{g}_{ij}^{\text{eCS2}}=&\frac{1}{1-\eta
}+\frac{3}{2}\frac{\eta(1-\eta/3) }{(1-\eta )^{2}}\frac{\sigma_i\sigma_j}{\sigma_{ij}}
\frac{\muM_2}{\muM_3}\nonumber\\
&+\frac{\eta^{2}(1-\eta/2)}{(1-\eta)^{3}}\left(\frac{\sigma_i\sigma_j}{\sigma_{ij}}
\frac{\muM_2}{\muM_3}\right)^{2}.
\label{15eCS2}
\end{align}

Equations \eqref{10.36} and \eqref{15SPT}--\eqref{15eCS2} have in common the fact that, at a given packing fraction $\eta$, the dependence of $\bar{g}_{ij}$ on
the sets of diameters $\{\sigma _{k}\}$ and mole fractions $\{x_{k}\}$ takes place \textit{only} through the scaled quantity
\begin{equation}
z_{ij}\equiv \frac{\sigma _{i}\sigma_{j}}{\sigma
_{ij}}\frac{\muM_{2}}{\muM_3}.
\label{zij}
\end{equation}
Thus, at a fixed packing fraction, the contact values of different pairs and different mixtures should collapse onto a common curve when plotted vs $z_{ij}$.
Figure \ref{g_vs_z} shows that this collapse property is reasonably well supported by  MC data for the six pair contact values of each one of five ternary mixtures at $\eta=0.49$.\cite{MMYSH02} The linear function \eqref{10.36} and the three quadratic functions \eqref{15SPT}--\eqref{15eCS2} are also plotted.
Comparison with the simulation data indicates that Eq.\ \eqref{15eCS2} performs generally better than Eq.\ \eqref{15BGHLL}, especially as $z_{ij}$ increases

Clearly, the use of \emph{any} of these approximate contact values in the RFA approach will lead to (almost completely) analytical  results for the structural properties of HS mixtures with the further asset that there will be full thermodynamic compatibility between the virial and compressibility routes. As an extra bonus, the (usually hard) numerical labor involved in the determination of the structural properties of these systems by solving integral equations is eliminated.

Figure \ref{ternary} presents a comparison between the results of the RFA method  with the PY theory and simulation data\cite{MMYSH02} for the like-like RDF of a ternary mixture.  In the case of the RFA, we have used the eCS2 contact values and the corresponding isothermal susceptibility. The improvement of the RFA over the PY
prediction, particularly in the region near contact, is noticeable. Although the RFA accounts nicely for the observed oscillations, it
seems to somewhat overestimate the depth of the first minimum.

A recent comparison with MD simulations\cite{PBYSH20} shows that the RFA is also rather accurate in predicting the DCF $c_{ij}(r)$ and the asymptotic large-$r$ behavior of the RDF $g_{ij}(r)$ for HS mixtures, including the so-called structural crossover transition.\cite{ELHH94,SPTEP16}

\section{Mappings of the EOS of the Single-Component HS Fluid onto the EOS of the HS Fluid Mixture}
\label{sec4}

Obviously, it is, in principle, simpler to obtain the compressibility factor of the one-component fluid, $Z_\pure$,\footnote{As already done in Eq.\ \eqref{z2a2X2X3},  the subscript $\pure$ is used in Sec.\ \ref{sec4} to refer to single-component quantities} than that of a multicomponent mixture, $Z$. On the other hand, it is well known that a certain degree of polydispersity is enough to prevent crystallization in HS fluids. In this section we present  the ideas we have followed to link the multicomponent compressibility factor, $Z$, and excess free energy per particle, $a^\ex$, to their one-component counterparts $Z_\pure$ and $a_\pure^\ex$. Among other things, this has allowed us to to examine (at least partially) the metastable fluid region of the monocomponent fluid.

\subsection{The e1 approximation}

Of course, a straightforward way to obtain approximate expressions for the EOS of the mixture in terms of the EOS of the single-component fluid is simply by proposing or deriving approximate expressions for the contact values $\bar{g}_{ij}$ in terms of $\bar{g}_\pure$. We have already followed this route and the outcome is briefly described below. More details may be found in Refs.\ \onlinecite{HYS08,S16} and references therein.

The basic assumption here is  that, given a certain packing fraction $\eta$, the whole dependence of $\bar{g}_{ij}$ on the composition $\{\sigma_{k},x_{k}\}$ and number of components $\Nc$ of the mixture occurs through the scaled quantity $z_{ij}$ defined in Eq.\ \eqref{zij}. More specifically,
\begin{equation}
\bar{g}_{ij}=\GG(\eta,z_{ij}),
\label{5}
\end{equation}
where the function $\GG(\eta,z)$ is \textit{universal}, {i.e.}, it is a \emph{common} function for all the pairs $(i,j)$, regardless
of the {composition and} number of components of the mixture.

As mentioned in Sec.\ \ref{sec3}, the PY, SPT,  BGHLL, and eCS2 contact values belong to the class of approximations \eqref{5}. Now, however, we want $\bar{g}_\pure$ to remain free, in contrast to what happens with Eqs.\ \eqref{10.36} and \eqref{15SPT}--\eqref{15eCS2}. This imposes the constraint
\begin{equation}
\label{single}
\GG(\eta,1)=\bar{g}_\pure.
\end{equation}
This means that the system is indistinguishable from a single-component one if all the sizes are identical ($\sigma_i=\sigma$ and $z_{ij}=1$).

Next, we consider the limit in which one of the species, say $i=0$, is made of
point particles, {i.e.}, $\sigma _{0}\rightarrow 0$. In that case,
$\bar{g}_{00}$ takes the ideal-gas value, except that one has
to take into account that the available volume for the point particles is not $V$ but $V(1-\eta)$.
Thus,
\begin{equation}
\lim_{\sigma_{0}\rightarrow
0}\bar{g}_{00}=\frac{1}{1-\eta}.
\label{2}
\end{equation}
Since $ z_{00}\rightarrow 0$ in the limit $\sigma
_{0}\rightarrow 0$, insertion of Eq.~(\ref{2}) into (\ref{5}) yields
\begin{equation}
\GG(\eta ,0)=\frac{1}{1-\eta }.
\label{6}
\end{equation}

As the simplest approximation of the class \eqref{5},\cite{SYH99} one may assume  a \emph{linear}
dependence of $\GG$ on $z$ that satisfies the basic requirements
(\ref{single}) and (\ref{6}), namely,
\begin{equation}
\GG(\eta,z)=\frac{1}{1-\eta }+\left( \bar{g}_{\pure}-\frac{1}{1-\eta }
\right) z.
\label{9}
\end{equation}
Inserting this into Eq.\ \eqref{5}, one has
\begin{equation}
\bar{g}_{ij}^{\text{e1}}=\frac{1}{1-\eta }+\left(
\bar{g}_{\pure}-\frac{1}{1-\eta
}\right)\frac{\sigma_i\sigma_j}{\sigma_{ij}}\frac{\muM_{2}}{\muM_3}.
\label{gije1}
\end{equation}
Here, the label ``e1'' is meant to indicate that (i) the contact
values used are an \emph{extension} of the single component contact
value $\bar{g}_{\pure}$ and that (ii) $\GG(\eta,z)$ is a \emph{linear}
polynomial in $z$.
When Eq.\ \eqref{gije1} is inserted into Eq.\ \eqref{ec1}, one easily obtains
\begin{align}
 Z_\text{e1}(\eta )=&1+\left(1+\frac{\mt}{\mth}-2\frac{\mt^3}{\mth^2}\right)\frac{\eta}{1-\eta}\nonumber\\
 &+\frac{1}{2}\left(\frac{\mt}{\mth}+\frac{\mt^3}{\mth^2}\right)\left[ Z_{\pure}(\eta )-1\right].
 \label{Ze1}
\end{align}

The proposal (\ref{gije1}) is rather crude and does not produce
especially accurate results for $\bar{g}_{ij}$. In fact, as Fig.\ \ref{g_vs_z} shows, the contact values clearly deviate from a linear dependence on $z_{ij}$. Nevertheless, the EOS \eqref{Ze1} exhibits an excellent
agreement with simulations, provided
that an accurate function $Z_{\pure}(\eta)$ is used as input.\cite{SYH99,MV99,SYH01,GAH01,HYS02}.

It is interesting to point out that from Eq.\ (\ref{Ze1}), one may write the reduced virial
coefficients of the mixture in terms of the reduced virial coefficients of the single-component fluid $b_n\equiv  B_{n,\pure}/(\frac{\pi}{6} \sigma^3)^{n-1}$. The result is
\begin{equation}
\bar{B}_n^{\text{e1}}=
1+\frac{\mt}{\mth}-2\frac{\mt^2}{\mth^3}
+\frac{b_n}{2}\left(\frac{\mt}{\mth}+\frac{\mt^2}{\mth^3}\right),\quad n\geq 2.
\label{Virial}
\end{equation}
Taking into account that $b_2=4$ and $b_3=10$, and comparing with Eq.\ \eqref{Second&Thirdvir}, we can see that the \emph{exact} second and third virial coefficients are recovered.
In general, however, $\bar{B}_n^{\text{e1}}$ with $n\geq 4$ are only approximate.

By eliminating $\mt/\mth$ and $\mt^2/\mth^3$ in favor of $\bar{B}_2$ and  $\bar{B}_3$, Eq.\  \eqref{Ze1} can be rewritten as
\begin{equation}
Z_\text{e1}(\eta)=1+\frac{5\bar{B}_2-2\bar{B}_3}{3}\frac{\eta}{1-\eta}+\frac{\bar{B}_3-\bar{B}_2}{6}\left[Z_\pure(\eta)-1\right].
\label{Ze1bis}
\end{equation}
This allows us to interpret that, in the e1 approximation, the excess compressibility factor $Z(\eta)-1$ is  just a linear combination of $\eta/(1-\eta)$ and $Z_\pure(\eta)-1$ with density-independent coefficients such that the exact second and third virial coefficients are retained.\cite{SYHO17,HSY20}

From Eq.\ \eqref{Zfroma} one can obtain the excess free energy per particle as [see first step in Eq.\ \eqref{fromZtomu}]
\begin{equation}
\betaT a_\text{e1}^\ex(\eta)=-\frac{5\bar{B}_2-2\bar{B}_3}{3}\ln(1-\eta)+\frac{\bar{B}_3-\bar{B}_2}{6}\betaT a^\ex_\pure(\eta).
\label{aexe1}
\end{equation}
Note that the e1 excess Helmholtz free energy per particle is  expressed in terms of $\eta$ and the first three size moments only. In general,  $a^\ex$ is said to have a \emph{truncatable} structure if it depends on the size distribution  \emph{only} through a \emph{finite} number of moments.\cite{GKM82,SWC01,S02}

\subsection{The e2 approximation}
\label{sec:e2}

The second approximation, labeled ``e2,'' similarly indicates that (i) the resulting contact values $\bar{g}_{ij}$ represent an \emph{extension} of the single-component contact value $\bar{g}_\pure$ and that (ii) the universal function ${\GG}(\eta, z)$ is a \emph{quadratic} polynomial in $z$.
We keep the basic requirements \eqref{single} and \eqref{6}, but now we need an extra condition.

To that end, let us consider an $(\Nc+1)$-component mixture where one of the species (here denoted as $i=0$) is made of a single particle ($x_0\to 0$) of infinite diameter ($\sigma_0\to\infty$), which thus acts as a \emph{wall}.\cite{HMLC96,HBCW98,RDA01} The contact value $\bar{g}_{wj}$ of the correlation function  of a sphere of
diameter $\sigma_j$ with the wall, namely,
\begin{equation}
\bar{g}_{wj}=\lim_{{\sigma_0\to\infty},{
x_{0}\sigma_{0}^3\rightarrow 0}} \bar{g}_{0j},
\label{3p}
\end{equation}
gives the ratio between the density of
particles of species $j$ adjacent to the wall and the density of
those particles far away from the wall. The sum rule connecting the
pressure of the fluid and the above contact values is\cite{E90}
\begin{equation}
1+
\frac{4 \eta}{\muM_3} \sum_{i,j=1}^{\Nc}x_i x_j\sigma_{ij}^3 \bar{g}_{ij}=\sum_{j=1}^{\Nc}  x_j \bar{g}_{wj}.
\label{4p}
\end{equation}
This condition means that, when the mixture is in contact with a hard wall, the state of equilibrium
imposes that the pressure evaluated near the wall by considering the
impacts of the particles with the wall must be the same as the pressure in the bulk
evaluated from the particle--particle collisions. This consistency
condition is especially important if one is interested in deriving
accurate expressions for the contact values of the particle--wall
correlation functions.

In terms of the universality ansatz \eqref{5}, Eq.\ \eqref{4p} reads
\begin{equation}
1+
\frac{4 \eta}{\muM_3} \sum_{i,j=1}^{\Nc}x_i x_j\sigma_{ij}^3 \GG(\eta,z_{ij})=\sum_{j=1}^{\Nc}  x_j \GG(\eta,z_{wj}),
\label{4pG}
\end{equation}
where
\begin{equation}
\label{zwj}
z_{wj}\equiv 2\sigma_j\frac{\mt}{\mth}.
\end{equation}
In the special case of a pure fluid ($\Nc=1$) plus a wall, Eqs.\ \eqref{4p} and \eqref{4pG} reduce to
\begin{subequations}
\begin{equation}
1+
{4 \eta} \bar{g}_{\pure}= \bar{g}_{w},
\label{4}
\end{equation}
\begin{equation}
1+
{4 \eta} \GG(\eta,1)=\GG(\eta,2),
\label{8}
\end{equation}
\end{subequations}
respectively.
Thus, Eqs.\ \eqref{single}, (\ref{6}), and (\ref{8}) provide  complete
information on the function $\GG$ at $z=1$, $z=0$, and $z=2$,
respectively, in terms of the contact value $\bar{g}_{\pure}$ of the
single-component RDF.

Therefore, the explicit expression for the contact values in the e2 approximation reads\cite{SYH02}
\begin{align}
\bar{g}_{ij}^{\text{e2}}=&\frac{1}{1-\eta }+\left[
2(1-\eta )\bar{g}_{\pure}-\frac{2-\eta/2}{1-\eta
}\right]\frac{\sigma_i\sigma_j}{\sigma_{ij}}\frac{\muM_2}{\muM_3}\nonumber\\
&+\left[\frac{1-\eta/2}{1-\eta }-(1-2\eta
)\bar{g}_{\pure}\right]\left(\frac{\sigma_i\sigma_j}{\sigma_{ij}}\frac{\muM_{2}}{\muM_3}\right)^2.
\label{gije2}
\end{align}
The eCS2 proposal \eqref{15eCS2} is obtained from Eq.\ \eqref{gije2} with the CS choice $\bar{g}_\pure=\bar{g}_\cs=(1-\eta/2)/(1+\eta)^3$. As seen in Fig.\ \ref{g_vs_z}, it provides a very good account of simulation results.
It is interesting to note that  if
one considers a binary mixture in the infinite solute dilution
limit, namely,  $x_1 \rightarrow 0$, so that $z_{12} \rightarrow
2/(1+\sigma_2/\sigma_1)$, Eq.~(\ref{15eCS2}) yields the same result for
$\bar{g}_{12}$ as the one proposed by Matyushov and Ladanyi\cite{ML97} for this  quantity on the basis of exact geometrical
relations. However, the extension that the same authors propose when
there is a non-vanishing solute concentration, {{i.e.,}} for
$x_1\neq 0$, is different from Eq.\ (\ref{15eCS2}).

Insertion of Eq.\ \eqref{gije2} into Eq.\ \eqref{ec1} yields
\begin{equation}
 Z_\text{e2}(\eta )=\frac{1}{1-\eta }+
 \left[ \frac{\mt}{\mth}\left(1-\eta\right) +\frac{
 \mt^{2}}{\mth^3 }\eta \right]\left[ Z_{\pure}(\eta )-\frac{1}{1-\eta
 }\right].
 \label{Ze2}
\end{equation}
The associated (reduced) virial coefficients are
\begin{equation}
\label{Virial_e2}
\bar{B}_n^{\text{e2}}=1-\frac{\mt^3}{\mth^2}+b_n\frac{\mt}{\mth}+b_{n-1}
\left(\frac{\mt^2}{\mth^3}-\frac{\mt}{\mth}\right),\quad n\geq 2.
\end{equation}
It may be readily checked that $\bar{B}_2^{\text{e2}}=\bar{B}_2$ and $\bar{B}_3^{\text{e2}}=\bar{B}_3$, and so, the exact second and third virial coefficients are also recovered in this approximation. Taking that into account, Eq.\ \eqref{Ze2} can be rewritten as
\begin{align}
Z_\text{e2}(\eta )=&\frac{1}{1-\eta }+\left( \frac{\bar{B}_2-1}{3} +\eta \frac{\bar{B}_3-3\bar{B}_2+2}{3}\right)
\nonumber\\&\times
\left[ Z_{\pure}(\eta )-\frac{1}{1-\eta }\right] .
 \label{Ze2bis}
\end{align}

Equation \eqref{Ze2bis} has an interpretation different from that of Eq.\ \eqref{Ze1bis}. First, note that the ratio $\eta/(1-\eta)$ represents a \emph{rescaled} packing fraction, i.e., the ratio between the volume $V_{\text{occ}}\equiv N\frac{\pi}{6}\muM_3$ occupied by the spheres and the remaining \emph{void} volume $V_{\text{void}}\equiv V-V_{\text{occ}}$.  Next, we can realize that $\eta Z(\eta)-\eta/(1-\eta)$ represents a (reduced) \emph{modified} excess pressure with respect to a \emph{modified} ideal-gas value corresponding to the void volume $V_{\text{void}}$, namely,
\begin{equation}
\Delta\widetilde{p}(\eta)\equiv\frac{\pi}{6}\muM_3\left(\betaT p-\frac{N}{V_{\text{void}}}\right)=\eta \left[Z(\eta)-\frac{1}{1-\eta}\right].
\label{X18}
\end{equation}
To avoid confusion with the conventional  (reduced) excess pressure $\eta \left[Z(\eta)-1\right]$,
we will refer to the quantity $\Delta\widetilde{p}$ as the (reduced) \emph{surplus} pressure. The associated surplus free energy per particle, $\Delta a$, is defined as
\begin{equation}
\label{asp}
\betaT \Delta a(\eta)=\betaT a^\ex(\eta)+\ln(1-\eta).
\end{equation}
Thus, according to Eq.\ \eqref{Ze2bis}, the surplus pressure of the mixture, $\Delta\widetilde{p}(\eta)$, is equal to that of the single-component fluid at the same packing fraction, $\Delta\widetilde{p}_\pure(\eta)$, multiplied by a linear function of density whose coefficients are determined by requiring agreement with the exact  second and third virial coefficients.

From the first step in Eq.\ \eqref{fromZtomu} we can obtain the excess free energy per particle as
\begin{align}
\label{aexe2}
\betaT a_{\text{e2}}^\ex(\eta)=&-\frac{2+2\bar{B}_2-\bar{B}_3}{3}\ln(1-\eta)+\frac{\bar{B}_2-1}{3}\betaT a^\ex_\pure(\eta)\nonumber\\
&+\frac{\bar{B}_3-3\bar{B}_2+2}{3}\eta\int_0^1 dt\,Z_\pure(\eta t).
\end{align}
In contrast to the e1 approximation [Eq.\ \eqref{aexe1}], now, the free energy of the mixture depends on the properties of the single-component fluid not only through $a_\pure^\ex$ but also through a density integral of the compressibility factor $Z_\pure$.

If the CS EOS is chosen for the single-component fluid, i.e., $Z_\pure(\eta)=Z_\cs(\eta)$, Eqs.\ \eqref{Ze1bis} and \eqref{Ze2bis} yield the extensions $Z_{\text{eCS1}}(\eta)$ and $Z_{\text{eCS2}}(\eta)$, respectively. One can check that, for a given mixture composition,  $Z_{\text{eCS1}}(\eta )>Z_{\text{BMCSL}}(\eta)>Z_{\text{eCS2}}(\eta )$. Since simulation data indicate that the
BMCSL EOS tends to underestimate the compressibility factor, it
turns out that the
performance of $Z_{\text{eCS1}}$ is, {paradoxically}, better than
that of $Z_{\text{eCS2}}$,\cite{SYH02} despite the fact that the
underlying linear approximation for the contact values is much less
accurate than the quadratic approximation. This shows that a rather
crude approximation such as Eq.~(\ref{gije1}) may lead to an
extremely good EOS.\cite{SYH99,SYH01,GAH01,HYS02} Interestingly,  $Z_{\text{eCS1}}$ was independently derived as the
second-order approximation of a Fundamental Measure Theory (FMT) for the
HS fluid by Hansen-Goos and Roth.\cite{HR06a}

\subsection{The e3 approximation}
\label{sec:e3}
Apart from Eqs.\ \eqref{2} and \eqref{4}, there exist extra consistency conditions that
are not necessarily satisfied by \eqref{gije2}.\cite{R88,HMLC96,HC98,HC98b,HC00,HBCW98,ML97,BS00,H94,V98,THM99}
In particular, it fails to fulfill Eq.\ \eqref{4p} if a true mixture ($\Nc\geq 2$) is in contact with a wall.
On the other hand, if $\GG(\eta,z)$ is assumed to be a cubic function of $z$ with suitable density-dependent coefficients, then Eqs. \eqref{4p} and \eqref{4pG} are satisfied with independence of the composition of the mixture. The resulting ``e3'' approximation is\cite{SYH05,HYS06,HYS08}
\begin{align}
\bar{g}_{ij}^\text{e3}=&\frac{1}{1-\eta}+\frac{3 \eta}{2
\left(1-\eta\right)^2}\frac{\sigma_i
\sigma_j}{\sigma_{ij}}\frac{\muM_2}{\muM_3}+\Bigg[(2-\eta)\bar{g}_{\pure}
\nonumber\\&
-\frac{2+\eta^2/4}{\left(1-\eta\right)^2}\Bigg]
\left(\frac{\sigma_i
\sigma_j}{\sigma_{ij}}\frac{\muM_2}{\muM_3}\right)^2+(1-\eta)
\left(\bar{g}_{\text{SPT}}-\bar{g}_{\pure}\right)
\nonumber\\&\times
\left(\frac{\sigma_i\sigma_j}{\sigma_{ij}}\frac{\muM_2}{\muM_3}\right)^3,
\label{e3}
\end{align}
where
\begin{equation}
\bar{g}_{\text{SPT}}=\frac{1-\eta/2+\eta^2/4}{(1-\eta)^3}
\end{equation}
is the SPT contact value for a single fluid. In fact, the
choice $\bar{g}_\pure=\bar{g}_{\text{SPT}}$ makes the e3 approximation
become the same as the e2 approximation, both reducing to the SPT for
mixtures [Eq.\ \eqref{15SPT}].

Insertion of $\bar{g}_\pure=\bar{g}_\cs$ in Eq.\ \eqref{e3} yields the extension $\bar{g}_{ij}^{\text{eCS3}}$.
Comparison with computer simulations shows\cite{SYH05,HYS06} that the eCS3 approximation gives better predictions than the eCS2 approximation for the particle--wall contact
values, while for
the particle--particle contact values both the eCS2 and eCS3 are of
comparable accuracy.

From Eq.\ \eqref{e3}, the compressibility factor, the virial coefficients, and the excess free energy per particle may be  obtained as
 \begin{subequations}
\begin{align}
Z_\text{e3}(\eta)=&\frac{1}{1-\eta}+\left(\frac{\mt}{\mth}-\frac{\mt^3}{\mth^2}\right)\frac{3
\eta}{
\left(1-\eta\right)^2}\nonumber\\
&+\frac{\mt^3}{\mth^2}\left[Z_\pure(\eta)-\frac{1}{1-\eta}\right]\nonumber\\
=&\frac{1}{1-\eta}+\left(3\bar{B}_2-\bar{B}_3-2\right)\frac{
\eta}{
\left(1-\eta\right)^2}\nonumber\\
&+\frac{\bar{B}_3-2\bar{B}_2+1}{3}\left[Z_\pure(\eta)-\frac{1}{1-\eta}\right],
\label{ZZ}
\end{align}
\begin{equation}
\bar{B}_n^{\text{e3}}=1-(3n-2)\frac{\mt^3}{\mth^2}+3(n-1)\frac{\mt}{\mth}
+\frac{\mt^3}{\mth^2}b_n,\quad n\geq 2,
\end{equation}
\begin{align}
\betaT a^\ex(\eta)=&-\frac{2+2\bar{B}_2-\bar{B}_3}{3}\ln(1-\eta)+\left(3\bar{B}_2-\bar{B}_3-2\right)\nonumber\\
&\times\frac{
\eta}{1-\eta}+\frac{\bar{B}_3-2\bar{B}_2+1}{3}\betaT a_\pure^\ex(\eta).
\end{align}
 \end{subequations}
According to Eq.\ \eqref{ZZ} the surplus pressure $\Delta\widetilde{p}(\eta)$ is expressed in the e3 approximation as a linear combination of $\Delta\widetilde{p}_\pure(\eta)$ and $\eta^2/(1-\eta)^2$ with density-independent coefficients ensuring consistency with the second and third virial coefficients.

It is worthwhile noting that the PY EOS for mixtures derived from the virial and compressibility routes are recovered from Eq.\ \eqref{ZZ} by choosing the corresponding PY one-component compressibility factors. As a consequence, the choice $Z_\pure(\eta)=Z_\cs(\eta)$ leads to the BMCSL EOS, i.e., $Z_{\text{eCS3}}(\eta)=Z_{\text{BMCSL}}(\eta)$, even though $\bar{g}_{ij}^{\text{eCS3}}\neq \bar{g}_{ij}^{\text{BGHLL}}$. This illustrates the possibility that different approximations for the contact values may yield a common EOS.

Summarizing the performance of the three approximations e1--e3, one can say that, provided an accurate single-component EOS is used as input, the best multicomponent EOS is obtained from the e1 approximation, followed by the e3 approximation. In contrast, the e1 contact values are too simplistic, while the e2 and e3 values are more accurate than the BGHLL ones, the e3 approximation being especially reliable for the particle--wall contact values.

\subsection{Class of consistent FMT. The sp approximation}
The three previous approaches are based on the universal ansatz \eqref{5} and follow the path $\bar{g}_{ij}\to Z\to a^\ex$. Now we are going to invert the path by starting from a FMT for the free energy and then obtaining the compressibility factor.
Moreover, while in the e1--e3 approximations the  one-component quantities  are evaluated at the same packing fraction as that of the mixture, we will not impose now this condition \emph{a priori}.
We will start from two basic requirements that are analogous to the limits in Eqs.\ \eqref{2} and \eqref{4p}.

\subsubsection{Two consistency conditions on the free energy}

Let us assume that, without modifying the volume $V$, we add $N_0=x_0N$ extra particles of diameter $\sigma_0$ to an $\Nc$-component mixture so that the augmented system has a number density $\rho'=\rho(1+x_0)$,  a set of mole fractions $\{x_0',x_1',x_2',\ldots,x_{\Nc}'\}$, where $x_i'=x_i/(1+x_0)$, and a packing fraction $\eta'=\eta+\rho x_0\frac{\pi}{6}\sigma_0^3$. The relationships between the original and augmented moments are
\begin{subequations}
\begin{equation}
\muM_n'=\frac{\muM_n+x_0\sigma_0^n}{1+x_0},
\end{equation}
\begin{equation}
\label{mnaug}
\mn'=\frac{\mn+x_0\sigma_0^n/\muM_1^n}{\left(1+x_0\sigma_0/\muM_1\right)^n}(1+x_0)^{n-1}.
\end{equation}
\end{subequations}

Now, if the extra particles have zero diameter ($\sigma_0\to 0$), it can be proved\cite{S12} that
\begin{align}
\label{X1b}
\lim_{\sigma_0\to 0}\betaT a^\ex\left(\eta;\left\{x_0',x_1',\ldots\right\}\right)=&\frac{\betaT a^\ex(\eta;\{x_1,\ldots\})}{1+x_0}\nonumber\\
&
-\frac{x_0}{1+x_0}\ln(1-\eta),
\end{align}
which holds for arbitrary $x_0>0$.

Next, we turn to another more stringent condition. Instead of taking the limit $\sigma_0\to 0$ for an arbitrary number $N_0$ of extra particles, we assume that $N_0\ll N$ (i.e., $x_0\to 0$) and  $\sigma_0\to\infty$ in such a way that $x_0\sigma_0^3/\eta\to 0$ (i.e., the few extra ``big'' particles occupy a negligible volume). In that case,\cite{RFHL60,R89,RELK02,R10}
\begin{equation}
\lim_{\sigma_0\to\infty,x_0\sigma_0^3\to 0}\frac{\mu_0^\ex(\eta';\{x_0',x_1',\ldots\})}{\frac{\pi}{6}\sigma_0^3}= p(\eta;\{x_1,\ldots\}).
\label{X3}
\end{equation}
This condition is related to the
reversible work needed to create a cavity large enough to
accommodate a particle of infinite diameter.
Using the thermodynamic relations \eqref{Zfroma} and \eqref{fromZtomu}, one can rewrite Eq.\ \eqref{X3} as
\begin{align}
\lim_{\sigma_0\to\infty,x_0\sigma_0^3\to 0}\frac{\partial  a^\ex(\{\rho_0,\rho_1,\ldots\})}{\frac{\pi}{6}\sigma_0^3\partial \rho_0}=&
\eta\frac{\partial  a^\ex(\eta;\{x_1,\ldots\})}{\partial \eta}
\nonumber\\
&+k_BT,
\label{X4}
\end{align}
where on the left-hand side the change in independent variables $(\eta';\{x_0', x_1',\ldots\})\to(\{\rho_0,\rho_1,\ldots\})$ has been carried out.

The exact conditions \eqref{X1b} and \eqref{X4} complement each other since the former accounts for the limit where one of the species is made of point particles, whereas the latter accounts for the opposite limit where a few particles have a very large size.

\subsubsection{Consistent  truncatable approximations}

Now we restrict ourselves to (approximate) free energies with a truncatable structure involving the first three moments, i.e.,
$ a^\ex(\eta;\{x_i\})\to a^\ex(\eta;\mt,\mth)$,  in the spirit of a FMT.\cite{R89,TCM08}
First, note from Eq.\ \eqref{mnaug} that in the limit $\sigma_0\to 0$, the reduced moments of the augmented system are $\mn'=(1+x_0)^{n-1}\mn$. Thus, Eq.\ \eqref{X1b} becomes
\begin{align}
\betaT a^\ex(\eta;\zeta \mt,\zeta^{2}\mth)+\ln(1-\eta)=&\frac{1}{\zeta}\left[\betaT a^\ex(\eta;\mt,\mth)\right.\nonumber\\
&\left.+\ln(1-\eta)\right],
\label{X6}
\end{align}
where $\zeta\equiv 1+x_0$ is arbitrary. This scaling property necessarily implies the functional form
\begin{equation}
\betaT a^\ex(\eta;\{x_i\})=\omega{\Ac\left(\eta,\lambda\right)}- \ln(1-\eta),
\label{X5}
\end{equation}
where
\begin{equation}
\label{omegalambda}
\omega\equiv \frac{1}{\mt},\quad \lambda\equiv \frac{\mth}{\mt^2},
\end{equation}
and $\Ac\left(\eta,\lambda\right)$ is so far an arbitrary unknown function.

We have not used the more stringent consistency condition \eqref{X4} yet. According to Eq.\ \eqref{X5}, Eq.\ \eqref{X4} yields
\begin{align}
\lim_{\sigma_0\to\infty,x_0\sigma_0^3\to 0}\frac{\partial}{\frac{\pi}{6}\sigma_0^3\partial \rho_0}\left[\frac{\Ac(\eta',\lambda')}{\mt'}-\ln(1-\eta')\right]\nonumber\\
=
\eta\frac{\partial}{\partial \eta}\left[\frac{\Ac(\eta,\lambda)}{\mt}-\ln(1-\eta)\right]+1.
\end{align}
This reduces to the first-order linear partial differential equation
\begin{equation}
\eta(1-\eta)\frac{\partial\Ac(\eta,\lambda)}{\partial\eta}+\lambda\frac{\partial\Ac(\eta,\lambda)}{\partial \lambda}=0,
\label{X12}
\end{equation}
where use has been made of the  properties
\begin{subequations}
\label{X8}
\begin{equation}
\frac{1}{\frac{\pi}{6}\sigma_0^3}\frac{\partial \eta'}{\partial \rho_0}=1,
\end{equation}
\begin{equation}
\lim_{\sigma_0\to\infty,x_0\sigma_0^3\to 0}\frac{1}{\frac{\pi}{6}\sigma_0^3}\frac{\partial \mt'}{\partial \rho_0}=0,
\end{equation}
\begin{equation}
\lim_{\sigma_0\to\infty,x_0\sigma_0^3\to 0}\frac{1}{\frac{\pi}{6}\sigma_0^3}\frac{\partial \lambda'}{\partial \rho_0}=\frac{\lambda}{\eta}.
\end{equation}
\end{subequations}
The general solution of Eq.\ \eqref{X12} is
\begin{equation}
\Ac(\eta,\lambda)=\Ac_0\left(\frac{\eta}{(1-\eta)\lambda}\right),
\label{X13}
\end{equation}
where $\Ac_0$ is a function of a single scaled variable. It is  determined by the one-component constraint
\begin{equation}
\Ac_0\left(\frac{\eta}{1-\eta}\right)=\betaT a_\pure^\ex(\eta)+\ln(1-\eta).
\label{X14}
\end{equation}

Combining Eqs.\ \eqref{X5}, \eqref{X13}, and \eqref{X14}, we finally obtain
\begin{align}
\label{CFMT}
\betaT a^\ex(\eta)=&\omega\left[\betaT a^\ex_\pure\left(\frac{\eta}{\eta+\lambda(1-\eta)}\right)\right.\nonumber\\
&\left.+\ln\frac{\lambda(1-\eta)}{\eta+\lambda(1-\eta)}\right]-\ln(1-\eta).
\end{align}
While Eq.\ \eqref{CFMT} can be extended to inhomogeneous situations as a FMT,\cite{S12c,L13,HMOR15} here we focus on homogeneous fluids.

\subsubsection{The sp approximation}
The class of free energies \eqref{CFMT} includes all the truncatable (first three moments) free energies that satisfy simultaneously the constraints \eqref{X1b} and \eqref{X4}.
Using the definition of surplus free energy [Eq.\ \eqref{asp}], we realize that Eq.\ \eqref{CFMT} implies that the surplus free energy of the mixture at a given packing fraction is just proportional to that of single-component fluid at a different \emph{effective} packing fraction, namely,
\begin{equation}
\Delta a_\spl(\eta)={\omega}\Delta a_\pure(\eff),
\label{17}
\end{equation}
where
\begin{equation}
\eta_{\text{eff}}=\frac{\eta}{\eta+\lambda(1-\eta)},\quad \frac{\eta_{\text{eff}}}{1-\eta_{\text{eff}}}=\frac{1}{\lambda}\frac{\eta}{1-\eta}.
\label{18a}
\end{equation}
A label ``sp'' has been introduced motivated by the simplicity of Eq.\ \eqref{17} when expressed in terms of surplus quantities.\cite{SYHOO14,S16,SYHO17,HSY20}
Note that the effective \emph{rescaled}  packing fraction of the single-component fluid is just the rescaled packing fraction of the mixture divided by $\lambda$.

The inequalities\cite{OL12} $\mth\geq \mt^2\geq \mt\geq 1$ imply $\lambda\geq 1\geq\omega$, so that $\eta\geq\eff$ but $\Delta a_\spl(\eta)\leq \Delta a_\pure(\eff)$. In terms of the second and third virial coefficients, the parameters  $\lambda$ and $\omega$ can be expressed as
\begin{equation}
\lambda=\frac{\bar{B}_2-1}{\bar{B}_3-2\bar{B}_2+1},\quad
\omega=\frac{(\bar{B}_2-1)^2}{3(\bar{B}_3-2\bar{B}_2+1)}.
\label{26}
\end{equation}
Thus, one can interpret the sp approximation by stating that the surplus free energy and rescaled packing fraction of the mixture are both proportional to their respective one-component counterparts, the proportionality constants being determined from the requirement that the exact second and third virial coefficients are kept. This simple interpretation allows for a successful straightforward extension of the sp approximation to hard disks\cite{SYHO17} and hard hyperspheres.\cite{HSY20}

\begin{figure}
  \includegraphics[width=77mm]{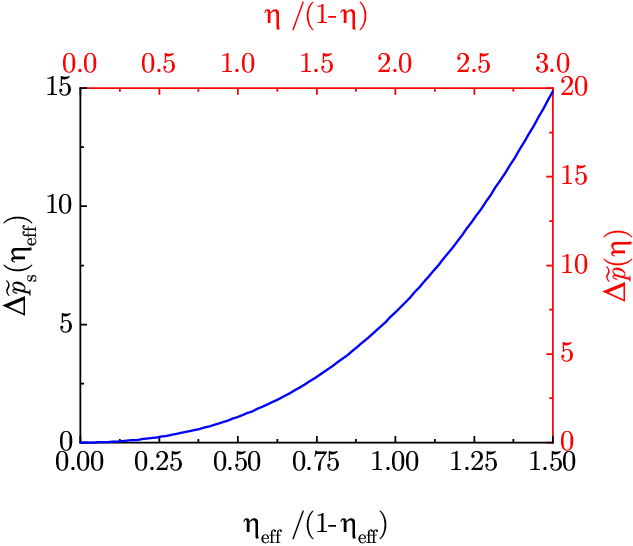}
\caption{(Color
  online) Graphical illustration of the mapping ``polydisperse mixture $\leftrightarrow$ pure fluid'' defined by Eqs.\ \protect\eqref{18a} and \protect\eqref{pspl}. A common curve represents the EOS of the pure fluid (left and bottom axes) and that of the mixture (top and right axes). In this particular example the mixture is characterized by $m_2=\frac{3}{2}$ and $m_3=\frac{9}{2}$, so that $\lambda=2$ and $\omega=\frac{2}{3}$. A change in the composition of the mixture would only be reflected by a rescaling of the top and right axes. In this graph the curve corresponds to the single-component CS surplus pressure. {Adapted from Fig.\ 1 of Ref.\ \onlinecite{SYHOO14}}.} \label{illustration}
\end{figure}

From the thermodynamic relation $\Delta \widetilde{p}=\eta^2\partial\betaT\Delta a/\partial \eta$, where the surplus pressure is defined in Eq.\ \eqref{X18}, Eq.\ \eqref{17} yields
\begin{equation}
\label{pspl}
\Delta \widetilde{p}_\spl(\eta)=\lambda \omega \Delta \widetilde{p}_\pure(\eff),
\end{equation}
where we have used $\partial \eff/\partial\eta=\lambda\eff^2/\eta^2$.
In terms of the compressibility factor, the mapping between the EOS of the monocomponent and multicomponent systems in the sp approximation becomes
\begin{equation}
Z_{\text{sp}}(\eta)=\frac{1}{1-\eta}+\lambda \omega\frac{\eff}{\eta}\left[Z_\pure(\eta_{\text{eff}})-\frac{1}{1-\eta_{\text{eff}}}\right].
\label{19}
\end{equation}
Since $\lambda\geq 1$ but $\omega\leq 1$, the product $\lambda \omega$ can be either larger or smaller than $1$, depending on the mixture composition. In the particular case of a binary mixture, it can be checked that $\lambda \omega>1$ only if the mole fraction of the small spheres is greater than a certain value, namely, $x_1>1/[1+(\sigma_1/\sigma_2)^{3/2}]>\frac{1}{2}$.

The mapping defined by Eqs.\ \eqref{18a} and \eqref{pspl} is illustrated by Fig.\ \ref{illustration}. A common curve describes both the multicomponent and the one-component systems, except that the top and right axes (corresponding to the mixture) are $\lambda$ and $\lambda\omega$ times, respectively, the bottom and left axes (corresponding to the pure fluid).

The e1--e3 and sp approximations can be formulated within a common framework in terms of the surplus pressure, as shown in the second column of Table \ref{e1e2e3sp}.
All of them share  the consistency with the exact second and third virial coefficients. On the other hand, as stressed before, while in the e1--e3 approximations the EOS of the mixture is related to that of the monocomponent fluid evaluated at the same packing fraction,  in the sp approximation the reference monocomponent fluid is evaluated at a different (effective) packing fraction $\eta_{\text{eff}}$.
Interestingly, if the monocomponent SPT surplus pressure
$\Delta\widetilde{p}_\pure(\eta)=3\eta^2/(1-\eta)^3$ is used, three of the four approximations (namely, e2, e3, and sp) yield the multicomponent SPT function $\Delta\widetilde{p}(\eta)=3\omega\eta^2[\eta+\lambda(1-\eta)]/\lambda^2(1-\eta)^3$. This proves the high degree of internal self-consistence of the SPT, even though it is not particularly accurate.
On the other hand, if a better one-component EOS is used as input (for instance, $Z_\pure=Z_{\text{CS}}$), one finds $Z_{\text{e1}}(\eta)>Z_{\text{sp}}(\eta)>Z_{\text{e3}}(\eta)>Z_{\text{e2}}(\eta)$.

\subsection{Comparison with computer simulations}

\begin{table*}
\caption{Surplus pressure of the multicomponent system, $\Delta\widetilde{p}$, expressed in terms of the surplus pressure of the single-component fluid, $\Delta\widetilde{p}_\pure$, according to the approximations e1--e3 and sp. The third column gives the single-component compressibility factor, $Z_\pure$, as inferred from that of the mixture, $Z$. The parameters $\lambda$ and $\omega$ are defined in Eq.\ \eqref{omegalambda}.
\label{e1e2e3sp}}
\begin{ruledtabular}
\begin{tabular}{lcc}
Approximation&$\Delta\widetilde{p}(\eta)$&$Z_\pure(\eta)$\\
\hline
e1&$\displaystyle{\frac{\omega(\lambda+1)}{2\lambda^2}\left[\Delta\widetilde{p}_\pure(\eta)+\frac{\lambda-1}{\lambda+1}\frac{3\eta^2}{1-\eta}\right]}$&$\displaystyle{\frac{1}{1-\eta}+\frac{2\lambda^2}{\omega(\lambda+1)}
\left[Z(\eta)-\frac{1}{1-\eta}\right]-\frac{\lambda-1}{\lambda+1}\frac{3\eta}{1-\eta}}$\\
e2&$\displaystyle{\frac{\omega}{\lambda^2}\Delta\widetilde{p}_\pure(\eta)\left[\eta+\lambda(1-\eta)\right]}$&$\displaystyle{\frac{1}{1-\eta}+\frac{\lambda^2}{\omega}
\frac{Z(\eta)-{1}/{(1-\eta)}}{\eta+\lambda(1-\eta)}}$\\
e3&$\displaystyle{\frac{\omega}{\lambda^2}\left[\Delta\widetilde{p}_\pure(\eta)+(\lambda-1)\frac{3\eta^2}{(1-\eta)^2}\right]}$&$\displaystyle{\frac{1}{1-\eta}+\frac{\lambda^2}{\omega}
\left[Z(\eta)-\frac{1}{1-\eta}\right]-(\lambda-1)\frac{3\eta}{(1-\eta)^2}}$\\
sp&$\lambda\omega \Delta\widetilde{p}_\pure(\eff),$\quad $\eff\equiv\displaystyle{\frac{\eta}{\eta+\lambda(1-\eta)}}$&$\displaystyle{\frac{1}{1-\eta}+\frac{\eta_{\text{m}}/\eta}{\lambda\omega}
\left[Z(\eta_{\text{m}})-\frac{1}{1-\eta_{\text{m}}}\right]}$,\quad $\displaystyle{\eta_{\text{m}}\equiv\frac{\eta}{\eta+(1-\eta)/\lambda}}$\\
\end{tabular}
\end{ruledtabular}
\end{table*}

\begin{figure*}
  \includegraphics[width=163mm]{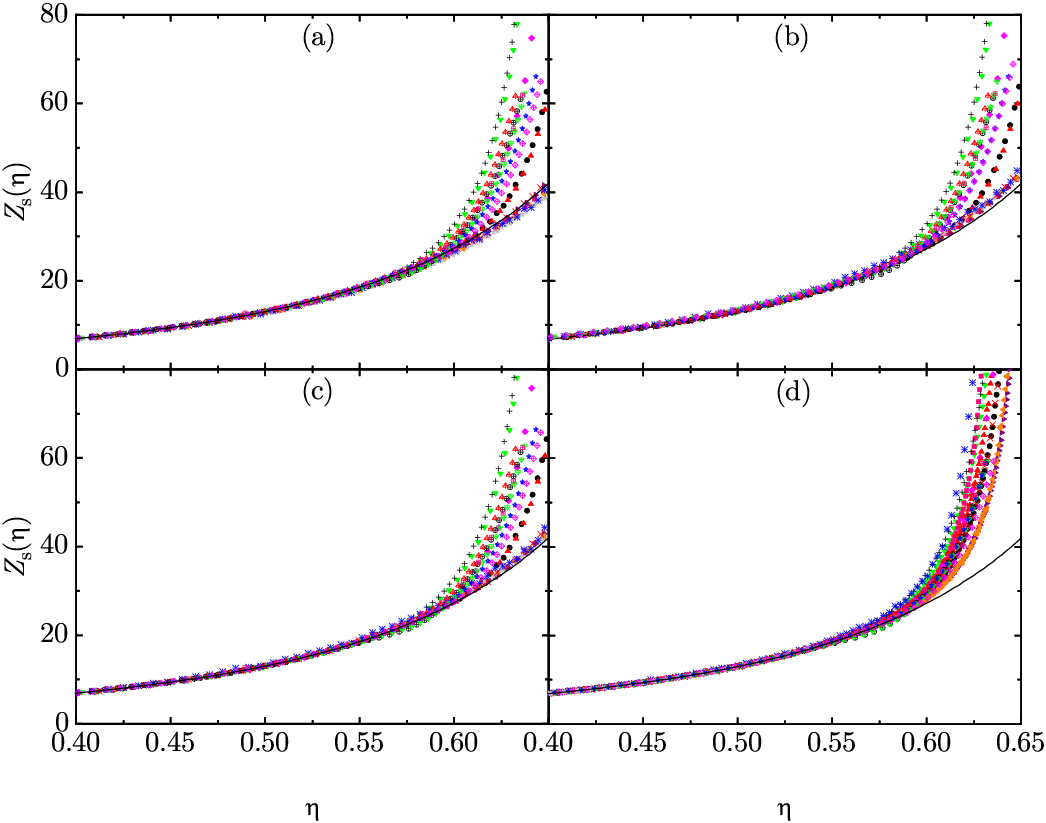}
\caption{Plot of the monocomponent  compressibility factor $Z_\pure(\eta)$, as inferred from simulation data for the mixtures described in Ref.\ \onlinecite{SYHOO14}, according to the (a) e1, (b) e2, (c) e3,  and (d) sp  prescriptions. The solid lines represent the CS EOS [Eq.\ \eqref{ZCS}].  {Panel (d) is adapted from Fig.\ 2(b) of Ref.\ \onlinecite{SYHOO14}}.}\label{fig_infer}
\end{figure*}

\begin{figure}
  \includegraphics[width=77mm]{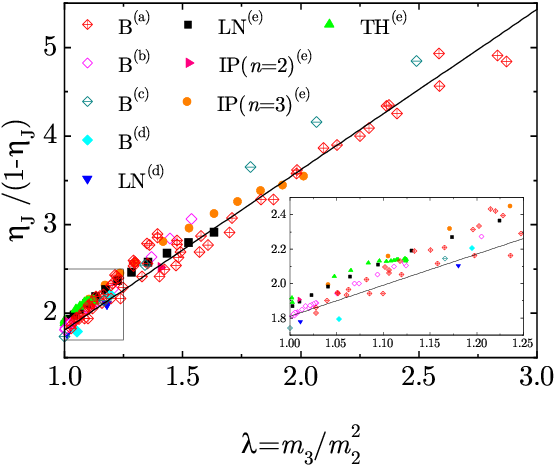}
\caption{Test of the prediction for the jamming packing fraction of polydisperse HS mixtures as given by Eq.\ \protect\eqref{jamming}, here represented by the straight solid line. Several classes of size distribution are considered: binary (B), log-normal (LN), truncated inverse-power (IP), and top-hat (TH). The superscripts indicate the sources: (a)  indicates experimental data  from Ref.\ \protect\onlinecite{YCW65}, (b) indicates simulation data from Ref.\ \protect\onlinecite{BCPZ09}, (c) indicates simulation data from Ref.\ \protect\onlinecite{HST13}, (d) indicates simulation data from Ref.\  \protect\onlinecite{DW14}, and (e) indicates simulation data from Refs.\ \protect\onlinecite{OL12,OL13,SYHOO14}. The inset is a magnifications of the framed region. {Adapted from Fig.\ 3(a) of Ref.\ \onlinecite{SYHOO14}}.
\label{etaJ}}
\end{figure}

It should be stressed that the proposals implied by Eqs.\ \eqref{Ze1}, \eqref{Ze2}, \eqref{ZZ}, and \eqref{19} may be interpreted in two directions. On the one hand, if $Z_\pure$ is known as a function of the packing fraction, then one can readily compute the compressibility factor of the mixture for any packing fraction and composition [$\eta_{\text{eff}}$ and $\eta$ being  related through Eq.\ \eqref{18a} in the case of $Z_{\text{sp}}$]; this is the standard view. On the other hand, if simulation data for the EOS of the mixture are available for different densities, size ratios, and mole fractions, Eqs.\ \eqref{Ze1}, \eqref{Ze2}, \eqref{ZZ}, and \eqref{19} can be used to \emph{infer} the compressibility factor of the monocomponent fluid.\cite{SYH11,SYHOO14} This is particularly important in the high-density region, where obtaining data from simulation may be accessible in the case of mixtures but either difficult or not feasible in the case of the monocomponent fluid, as that happens in the metastable fluid branch.\cite{SYHO17,SYHOO14} The third column of Table \ref{e1e2e3sp} provides the expressions of the single-component fluid compressibility factor, as inferred from that of the multicomponent system, for the four approximations discussed in this section.

In principle, simulation data for different mixtures would yield different inferred functions $Z_\pure(\eta)$. Thus, without having to use an externally imposed monocomponent EOS, the degree
of collapse of the mapping from mixture compressibility factors  onto a \emph{common} function $Z_\pure(\eta)$ is an efficient way of assessing the performance of Eqs.\ \eqref{Ze1}, \eqref{Ze2}, \eqref{ZZ}, and \eqref{19}. Let us test those mappings $Z\to Z_\pure$ by using simulation data corresponding to the mixtures described in Ref.\ \onlinecite{SYHOO14}, namely, $6$ binary mixtures and $11$ polydisperse fluids with three classes of continuous size distributions (top-hat, truncated inverse-power, and log-normal) and the explored values of the parameters $\lambda$ and $\omega$ being $1.01\lesssim\lambda\lesssim 1.63$ and $0.61\lesssim\omega\lesssim 0.99$.

As shown in Figure \ref{fig_infer}, the usefulness of the mappings is confirmed  by the nice collapse for all the points up to packing fractions below approximately the glass transition value $\eta_\text{g} \simeq 0.58$.\cite{PZ10,S98b} This includes  the metastable region beyond the freezing point ($\eta_\text{f}\simeq 0.492$), where simulation results for the monocomponent system are difficult to obtain.\cite{OB11} Beyond $\eta_\text{g}$, the collapse clearly fails in the e1--e3 inferences. On the other hand, the spread is significantly reduced in the case of the sp inference, although a certain degree of dispersion still remains, as expected from an
algorithm-dependent out-of-equilibrium glass branch.  This gives support to our expectation
that the relationship implied by Eq.\ (\ref{19}) (when extrapolated
to metastable states) might be useful for inferring the EOS of a metastable pure HS fluid from the knowledge of
the high-density behavior of polydisperse HS mixtures, which
is much more accessible than in the monodisperse case.\cite{OB11}

As a consequence of their ability to attain higher packing
fractions than the monocomponent system, the jamming packing
fraction $\eta_\text{J}$ of a polydisperse HS system is typically higher than the jamming packing
fraction (also called random close-packing fraction) $\eta_\text{J,\pure}$ of the one-component (monodisperse) system. Also, as that happens with $\eta_\text{J,\pure}$, the value of $\eta_\text{J}$ depends on the out-of-equilibrium compression protocol followed to
jam the system.\cite{PZ10,BCPZ09} Apart from that, since $\eta_\text{J}$ is a functional of the full size distribution function $\fx(\sigma)$, it differs widely from system to system without an apparent unifying framework. It obviously would be desirable  to have a way to characterize the whole distribution by a single ``dispersity''
parameter such that the actual values of $\eta_\text{J}$ for different polydisperse systems  would tend to approximately fall on a ``universal'' curve when plotted against such a dispersity parameter. This would provide  a tool to organize the apparently disconnected bunch of jamming packing values $\eta_\text{J}$ of different mixtures. We have shown that this is indeed possible.\cite{SYHOO14} What we have done is to assume that Eqs.\ (\ref{18a}) and (\ref{19})  are still valid (at least semiquantitatively) near the jamming point and take the value of $\lambda$ [see Eq.\ \eqref{omegalambda}] as the dispersity parameter (with the understanding that $\lambda = 1$ defines the monodisperse case). Since one must have $\lim_{\eta \to \eta_\text{J,\pure}} Z_\pure(\eta)=\infty$ and $\lim_{\eta \to \eta_\text{J}} Z_{\text{sp}}(\eta)=\infty$, Eq.\ (\ref{18a}) implies that $\eta_\text{J}$ is such that its associated effective one-component value of $\eta_\text{eff}$ coincides with $\eta_\text{J,\pure}$. Hence,
it follows that
\begin{equation}
\label{jamming}
\frac{\eta_\text{J}}{1-\eta_\text{J}}=\lambda\frac{\eta_\text{J,\pure}}{1-\eta_\text{J,\pure}},\quad \eta_\text{J}=\frac{\eta_\text{J,\pure}}{\eta_\text{J,\pure}+(1-\eta_\text{J,\pure})/\lambda}.
\end{equation}
Equation (\ref{jamming}) fulfills the  requirement posed above. First, all the details of the size distribution function $\fx(\sigma)$ are encapsulated in the moment ratio $\lambda=\mth/\mt^2$, which then plays the role of the sought dispersity parameter. Secondly, the functional form $\eta_\text{J}(\lambda)$ is quite simple: the occupied/void
volume ratio at jamming, i.e., ${\eta_\text{J}}/({1-\eta_\text{J}})$, is just proportional
to $\lambda$ with the slope being given by the monocomponent value.
Prediction (\ref{jamming}) is favorably tested against experimental\cite{YCW65} and simulation\cite{BCPZ09,OL12,OL13,HST13,DW14,SYHOO14} data in Fig.\ \ref{etaJ} taking $\eta_\text{J,\pure} = 0.644$.\cite{BCPZ09}

\section{Concluding Remarks}
\label{sec6}

In this paper we have attempted to provide a self-contained account of some of our (almost completely analytical) efforts over the past three decades in the study of the thermodynamic and structural properties of HS systems which have catered for, whenever possible, the consideration of mixtures with an arbitrary number of components (including polydisperse systems). These efforts include the RFA formalism, which is certainly an alternative methodology to the usual integral equation approach employed in liquid-state theory to obtain the structural properties of HS fluids.
Using this formalism, we have been able to obtain explicit analytical or semi-analytical results  for the RDF, the  DCF, the static structure
factor, and the bridge functions, in the end requiring as input
\emph{only} the contact value of the RDF of the single-component or multicomponent HS fluid.  One of the nice assets of the RFA is that, by construction, it eliminates the thermodynamic
consistency problem which is present in most of the integral
equation formulations for the computation of structural quantities. Moreover, this formalism has been amply shown to certainly improve the agreement between theory and computer simulation results in comparison,  say, to both the PY and HNC approximations. Very recently, for instance, we have also used the RFA to investigate the long-range behavior of the structural correlation functions, including the feature of structural crossover, in binary mixtures of additive HS.\cite{PBYSH20}

The RFA methodology can be applied beyond three-dimensional HS systems.
Those systems can be  classified into two categories: (i) those that admit an exact solution of the PY integral equation and (ii) those that are not exactly solvable within the PY approximation.
In the  first class of systems, the RFA method not only recovers, on the one hand, and improves, on the other hand, the PY solution for three-dimensional single-component or multicomponent HS fluids, but also does for hard hyperspheres\cite{RS07,RS11,RS11b}  and three-dimensional sticky-hard-sphere systems.\cite{YS93b,SYH98,YSH08}.
The application of RFA-like approaches to systems of the second class (i.e., those lacking an exact PY solution) includes the penetrable-sphere model,\cite{MS06,MYS07} the penetrable-square-well model,\cite{FGMS09} the square-well and square-shoulder potentials,\cite{YS94,AS01,LSYS05,YSH11,HYS16}, piecewise-constant potentials with several steps,\cite{SYH12,SYHBO13,HRYS18} nonadditive HS mictures,\cite{FS11,FS13,FS14}, and Janus particles with constrained orientations.\cite{MFGS13} In those cases,  the \emph{simplest} RFA is already quite accurate, generally improving on the (numerical) solution of the  PY approximation. Moreover, the RFA  is also directly applicable within the thermodynamic perturbation approach to the computation of the properties of real fluids.

The other area in which we have put attention here, where our work has also produced some interesting results, is the one pertaining to the EOS of HS mixtures. In this case, starting from relatively simple but otherwise reasonable assumptions whose merits may be judged \emph{a posteriori}, we have made use  of some exact consistency conditions to devise various approximate proposals that extend any given single-component EOS to  multicomponent fluid  mixtures with any composition and which are in  good agreement with simulation results. Some noteworthy aspects of the results that follow from
these developments were illustrated here through the prediction of the jamming packing-fraction of multicomponent systems as a simple function of a quantity involving the first three moments of the size distribution.
We should also mention that the ideas presented here in connection with the EOS for additive multicomponent HS mixtures have also been used for nonadditive systems and for hard-hypersphere systems  with arbitrary spatial dimensions.\cite{GAH01,SHY05,HYS08,SHY10,FSPHSY17,HSY20}

Finally, it should be clear that there are many facets of the
equilibrium and structural properties of  many other hard-core systems that may be studied following the ideas presented here but that up to now have not been considered. For instance, the generalization of the RFA approach to square-well or square-shoulder mixtures  and the surplus mapping of the single-component EOS to that of nonadditive HS mixtures  appear as  interesting challenges. We hope to address some of these
problems in the future and would be very much rewarded if some
others were taken up by researchers who might find these
developments also a valuable tool for their work.

\acknowledgments
A.S. and S.B.Y. acknowledge
financial support from the Spanish Agencia Estatal de Investigaci\'on
through Grant No.\ FIS2016-76359-P  and from the Junta de Extremadura
(Spain) through Grant No.\ GR18079, both partially financed
by the European Regional Development Fund.

\appendix

\section{On the functional forms of $\widetilde{h}(q)$ and $\widetilde{c}(q)$}
\label{appA}

In this appendix we present the expected functional forms of the correlation functions $\widetilde{h}(q)$ and $\widetilde{c}(q)$ in terms of an auxiliary function related to the Laplace transform $G(s)$ defined by Eq.\ \eqref{2.1G}.

For simplicity, we choose here $\sigma=1$ as the length unit. Let us define $\FF(s)\equiv\left[{2\pi \Psi(s)}\right]^{-1}$ and rewrite Eqs.\ \eqref{2.2} and \eqref{GGs} as
\begin{subequations}
\label{A3-A6}
\begin{equation}
\label{A3}
G(s)=s\frac{\FF(s) e^{-s}}{1+12\eta \FF(s) e^{-s}},
\end{equation}
\begin{equation}
\label{A6}
G(s)=s\sum_{n=1}^\infty (-12\eta)^{n-1}\left[F(s)\right]^n e^{-ns}.
\end{equation}
\end{subequations}
The small-$s$ behavior \eqref{2.4} translates into
\begin{equation}
\label{A4a}
\frac{e^s}{\FF(s)}=-12\eta+s^3+\mathcal{O}(s^5).
\end{equation}
On the other hand, the large-$s$  behavior is
\begin{subequations}
\begin{equation}
\label{A4b}
\FF(s)=\frac{\FF_{2}}{s^2}+\frac{\FF_{3}}{s^3}+\frac{\FF_{4}}{s^4}+\frac{\FF_{5}}{s^5}+\cdots,
\end{equation}
\begin{align}
\label{A5}
\FF_n=&\lim_{r\to 1^+}\frac{\partial^{n-2}}{\partial r^{n-2}}\left[r g(r)\right]\nonumber\\
=&(n-2)\bar{g}^{(n-3)}+\bar{g}^{(n-2)}.
\end{align}
\end{subequations}
This extends Eq.\ \eqref{2.3} to higher-order terms. In Eq.\ \eqref{A5}, $\bar{g}^{(n)}\equiv \lim_{r\to 1^+}\partial^n g(r)/\partial r^n$.

Let us now insert Eqs.\ \eqref{A3-A6} into Eq.\ \eqref{new2} to get
\begin{widetext}
\begin{subequations}
\begin{equation}
\label{Ahq}
\widetilde{h}(q)=-2\pi\frac{\FF(\imath q)e^{-\imath q}+\FF(-\imath q)e^{\imath q}+24\eta \FF(\imath q)\FF(-\imath q)}{1+12\eta \FF(\imath q)e^{-\imath q}+12\eta \FF(-\imath q)e^{\imath q}+144\eta^2\FF(\imath q)\FF(-\imath q)},
\end{equation}
\begin{equation}
\label{A9}
\widetilde{h}(q)=\sum_{n=1}^\infty \left[C^{(n)}(q)\cos(nq)+D^{(n)}(q)\sin(nq)\right].
\end{equation}
\end{subequations}
\end{widetext}
In Eq.\ \eqref{A9},
\begin{subequations}
\label{A10}
  \begin{equation}
\label{A10a}
C^{(n)}(q)=-2\pi(-12\eta)^{n-1}\left\{\left[\FF(\imath q)\right]^n+\left[\FF(-\imath q)\right]^n\right\},
\end{equation}
  \begin{equation}
  \label{A10b}
D^{(n)}(q)=2\pi(-12\eta)^{n-1}\imath\left\{\left[\FF(\imath q)\right]^n-\left[\FF(-\imath q)\right]^n\right\}.
\end{equation}
\end{subequations}
Taking into account the expansion \eqref{A4b}, one can write
\begin{subequations}
\label{A33}
\begin{align}
\left[\FF(s)\right]^n=&\frac{1}{s^{2n}}\left(\FF_{2}+\frac{\FF_{3}}{s}+\frac{\FF_{4}}{s^2}+\cdots\right)^n\nonumber\\
=&\frac{\FF_{2}^{n}}{s^{2n}}\left[1+\frac{n\FF_{3}}{\FF_{2}s}+n\frac{(n-1)\FF_{3}^2+2\FF_{2}\FF_{4}}{2\FF_{2}^2s^2}+\cdots\right],
\end{align}
\begin{align}
C^{(n)}(q)=&4\pi(12\eta)^{n-1}\frac{\FF_{2}^{n}}{q^{2n}}\nonumber\\
&\times\left[1-n\frac{(n-1)\FF_{3}^2+2\FF_{2}\FF_{4}}{2\FF_{2}^2q^2}+\mathcal{O}(q^{-4})\right],
\end{align}
\begin{equation}
D^{(n)}(q)=-4\pi(12\eta)^{n-1}\frac{\FF_{2}^{n}}{q^{2n}}
\left[\frac{n\FF_{3}}{\FF_{2}q}+\mathcal{O}(q^{-3})\right].
\end{equation}
\end{subequations}
Therefore, the large-$q$ structure of $C^{(n)}(q)$ and $D^{(n)}(q)$ is
\begin{subequations}
\label{A11}
  \begin{equation}
  \label{A11a}
C^{(n)}(q)=\sum_{m=n}^\infty \frac{C^{(n)}_m}{q^{2m}},
\end{equation}
  \begin{equation}
  \label{A11b}
D^{(n)}(q)=\sum_{m=n}^\infty \frac{D^{(n)}_m}{q^{2m+1}}.
\end{equation}
\end{subequations}
In particular,
\begin{subequations}
\label{A12}
  \begin{equation}
\label{A12a}
C^{(1)}_m=-4\pi (-1)^m\FF_{2m},
\end{equation}
  \begin{equation}
\label{A12b}
D^{(1)}_m=4\pi (-1)^m\FF_{2m+1},
\end{equation}
  \begin{equation}
\label{A12c}
C^{(n)}_n=4\pi (12\eta)^{n-1}\FF_{2}^n,
\end{equation}
  \begin{equation}
C^{(n)}_{n+1}=-2\pi (12\eta)^{n-1}n\FF_{2}^{n-2}\left[(n-1)\FF_3^2+2\FF_2\FF_4\right],
\end{equation}
  \begin{equation}
\label{A12d}
D^{(n)}_n=-4\pi (12\eta)^{n-1}n \FF_{2}^{n-1}\FF_{3}.
\end{equation}
\end{subequations}
It is worth noticing that the structure of Eqs.\ \eqref{A9} and \eqref{A11} agrees with that of Eq.\ (9) of Ref.\ \onlinecite{PBH17} only if the terms with $n>1$ in Eq.\ \eqref{A9} are ignored. On the other hand, the coefficients $C^{(1)}_1$, $C^{(1)}_2$, $D^{(1)}_1$, and $D^{(1)}_2$, as given by Eqs.\ \eqref{A5} and \eqref{A12}, agree with Eqs.\ (10)--(11) of Ref.\ \onlinecite{PBH17}.

Let us consider now the DCF. It can be obtained in Fourier space by insertion of Eq.\ \eqref{Ahq} into the OZ relation \eqref{OZFourier}. The result is
\begin{align}
\label{A9C}
\widetilde{c}(q)=&-{2\pi}\frac{\FF(\imath q)e^{-\imath q}+\FF(-\imath q)e^{\imath q}+24\eta \FF(\imath q)\FF(-\imath q)}{1-144\eta^2\FF(\imath q)\FF(-\imath q)}
\nonumber\\
=&\bar{C}^{(0)}(q)+ \bar{C}^{(1)}(q)\cos q+\bar{D}^{(1)}(q)\sin q,
\end{align}
 with
\begin{subequations}
\label{A22}
\begin{equation}
\bar{C}^{(0)}=-48\pi\eta\frac{ \FF(\imath q)\FF(-\imath q)}{1-144\eta^2 \FF(\imath q)\FF(-\imath q)},
\end{equation}
\begin{equation}
\label{A22b}
\bar{C}^{(1)}=-2\pi\frac{ \FF(\imath q)+\FF(-\imath q)}{1-144\eta^2 \FF(\imath q)\FF(-\imath q)},
\end{equation}
\begin{equation}
\bar{D}^{(1)}=2\pi\imath\frac{ \FF(\imath q)-\FF(-\imath q)}{1-144\eta^2 \FF(\imath q)\FF(-\imath q)}.
\end{equation}
\end{subequations}

Equations \eqref{A9} and \eqref{A9C} are formally exact and express the Fourier transforms $\widetilde{h}(q)$ and $\widetilde{c}(q)$ in terms of the (unknown) function $\FF(s)$ [see Eq.\ \eqref{A3}] evaluated at $s=\pm\imath q$. In the special case that $\FF(s)$ is an \emph{algebraic} function, all the singularities of $h(r)$ occur at $r=n$ and the only singularity of $c(r)$ takes place at $r=1$.

\section{Expressions for the amplitudes in Eq.\ \eqref{c(r)}}
\label{appB}
The explicit expressions for the amplitudes $\aK_{\pm}$, $\aK_{-1}$, $\aK_{0}$, $\aK_{1}$, $\aK_{3}$, and $\aK$ appearing in Eq.\ \eqref{c(r)} are
\begin{widetext}
\begin{subequations}
\begin{align}
 \aK_\pm=&\frac{ e^{\mp\kappa}}{ \alpha^2 (1 - \eta)^4
\kappa^6} \Bigl\{ 1+2(1+3 \alpha)\eta\pm \left[1 +
\frac{\eta}{2}+ \alpha (1 + 2 \eta )\right] \kappa
+
\frac{1 - \eta}{2}\left[\kappa^2 - \eta \left(12 +(\kappa \pm
6)\kappa\right)\right]\frac{L^\two }{\pi}\Big\}\nonumber\\
&\times\Bigl\{ 6 \eta
\left[1+2(1+3 \alpha)\eta\right] \pm 3 \eta \left[3\eta
- 2 \alpha (1-4 \eta)\right] \kappa - 3 \eta(1 + 2 \alpha)(1 -\eta)
\kappa^2 - \frac{(1 - \eta)^2}{2} \kappa^3 (\alpha \kappa\mp 1)\nonumber\\
 &+ 3\eta
(1 - \eta)  \left[\kappa^2 - \eta \left(12 +(\kappa \pm
6)\kappa\right)\right]\frac{L^\two}{\pi} \Big\},
\label{eca2}
\end{align}
\begin{equation}
\aK_{-1}=-\left[\frac{L^\two}{2 \pi \alpha} + \aK_+  e^{\kappa} +
\aK_-  e^{-\kappa}+ \aK_0 + \aK_1 + \aK_3\right],
\label{eca3}
\end{equation}
\begin{equation}
\aK_0=-\left[\frac{1+2\left(1+3 \alpha\right)\eta-6 \eta  \left( 1 -
\eta\right)L^\two/\pi }{\alpha \kappa \left(1 -\eta
\right)^2}\right]^2 ,
\label{eca4}
\end{equation}
\begin{align}
\aK_1=&\frac{6\eta}{\kappa^2}\aK_0+\frac{6 \eta}{\alpha^2
\kappa^2 \left(1 -\eta \right)^4}\Big\{\left[1+\frac{\eta}{2}+\alpha(1+
2\eta)\right]^2
\nonumber\\
& - \left(1 - \eta\right)\left[1 +\eta
  (7
+ \eta + 6 \alpha \left(2 + \eta \right))\right]\frac{L^\two}{\pi}\nonumber\\
& +3 \eta
\left(2 +\eta\right)(1-\eta)^2\frac{{L^\two}^2}{\pi^2}\Big\},
\label{eca5}
\end{align}
\begin{equation}
\aK_3=\frac{ \eta}{2}\aK_0,
\label{eca6}
\end{equation}
\begin{equation}
\aK=-\left(\aK_++\aK_-+\aK_{-1}\right).
\label{aprima}
\end{equation}
\end{subequations}
\end{widetext}
Here we have again taken $\sigma=1$ as the length unit.

\section*{DATA AVAILABILITY}

The data that support the findings of this study are available from the corresponding author upon reasonable request.

\section*{REFERENCES}

\bibliography{D:/Dropbox/Mis_Dropcumentos/bib_files/liquid}

\end{document}